\documentclass[12pt]{article}

\pdfoutput=1
\usepackage[latin9]{inputenc}
\usepackage[top=80pt,bottom=85pt,left=67pt,right=67pt]{geometry}
\usepackage{amssymb,amsmath}
\usepackage{xypic,subfigure}
\usepackage{graphicx,color}
\usepackage{bbm}
\usepackage{float}
\usepackage{cite}
\usepackage{tensor}
\usepackage{slashed}
\usepackage[debug,pageanchor=false]{hyperref}
\definecolor{link}{rgb}{.8,.15,.1}
\hypersetup{colorlinks=true,linkcolor=link,citecolor=link,urlcolor=link,linktocpage}

\makeatletter
\@addtoreset{equation}{section}
\makeatother

\newcommand{\beq}{\begin{equation}}
\newcommand{\eeq}{\end{equation}}
\newcommand{\bea}{\begin{eqnarray}}
\newcommand{\eea}{\end{eqnarray}}
\newcommand{\nn}{\nonumber}

\newcommand{\eq}{\begin{equation}}
\newcommand{\feq}{\end{equation}}
\newcommand{\eqn}{\begin{eqnarray}}
\newcommand{\feqn}{\end{eqnarray}}

\begin{document}

\begin{titlepage}

\begin{center}

\vskip .5in %.3in 
\noindent

{\Large \bf{Supersymmetric $\mathbb{WCP}^n$, AdS near horizons and orbifolds}}

\bigskip\medskip

Andrea Conti$^{a,b}$\footnote{contiandrea@uniovi.es}, Niall T. Macpherson$^{a}$\footnote{macphersonniall@uniovi.es}\\

\bigskip\medskip
{\small 

$a$: Department of Physics, University of Oviedo,
C/ Leopoldo Calvo Sotelo, 18, 33007 Oviedo, Asturias}

\medskip
{\small and}

\medskip
{\small 

Instituto Universitario de Ciencias y Tecnolog\'ias Espaciales de Asturias (ICTEA),\\
Calle de la Independencia 13, 33004 Oviedo, Spain}\\

\medskip
{\small

$b$: Blackett Laboratory, Imperial College,
Prince Consort Rd., London, SW7 2AZ, U.K }

\vskip 2cm 
%.6cm

     	{\bf Abstract }\\[2mm]
We construct and study the supersymmetry properties of the weighted projective spaces $\mathbb{WCP}^2$ and $\mathbb{WCP}^3$. These are topologically $\mathbb{CP}^n$ with $n+1$ orbifold singularities and as such are higher dimensional analogues of the ``spindle'' or $\mathbb{WCP}^1$. We use these to construct interesting supersymmetric orbifolds of canonical near horizon geometries of relevance to the AdS/CFT correspondence. Interestingly, for certain tunings of their integer weights, and unlike the spindle, $\mathbb{WCP}^{2}$ and $\mathbb{WCP}^{3}$ are compatible with supersymmetry beyond the realm of gauged supergravity. This allows one to construct interesting supersymmetric solutions in type II supergravity such as AdS$_5\times\mathbb{WCP}^{2}\times\text{S}^1$ and AdS$_4\times \mathbb{WCP}^3$ via duality, in which $\mathbb{WCP}^n$ appears without a connection fibred over it. We also leverage our results to construct a supersymmetric AdS$_3$ solution containing a topological $\mathbb{T}^{(1,1)}$ space with 4 orbifold singularities.  
     	\end{center}
     	\noindent

\noindent

\vfill
\eject

\end{titlepage}

\tableofcontents

\section{Introduction}
Orbifolds have a long history in string theory and related topics. Starting with \cite{Ferrero:2020laf} an orbifold that has played a prominent role in supergravity constructions in recent years is the spindle, which is the weighted projective space $\mathbb{WCP}^1_{[k_1,k_2]}$. This is a topological space with the behaviour of a 2-sphere with an $\mathbb{R}^2/\mathbb{Z}_{k_1}$ and $\mathbb{R}^2/\mathbb{Z}_{k_2}$ orbifold at the respective poles, a rational Euler characteristic of $\chi_{E}= \frac{k_1+k_2}{k_1k_2}$, and where usually\footnote{This assumption has been recently relaxed in \cite{Arav:2025jee}.} $\text{gcd}(k_1,k_2)=1$. The spindle requires a gauge field ${\cal A}$ to be consistent with supersymmetry which could appear as part of a circle fibration over $\mathbb{WCP}^1_{[k_1,k_2]}$, or in a gauged supergravity context a part of the spinorial supersymmetry conditions. In either cases the spindle can only preserve supersymmetry in one of two ways \cite{Ferrero:2021etw} such that the charge of $d{\cal A}$ takes one of two rational values
\beq
\frac{1}{2\pi}\int_{\text{WCP}^1}d{\cal A}=\frac{k_1+k_2}{k_1k_2},~~~~\frac{1}{2\pi}\int_{\text{WCP}^1}d{\cal A}=\frac{k_1-k_2}{k_1k_2},
\eeq  
these scenarios are referred to as a twist and an anti-twist respectively. 

Spindles are interesting in part because they provide an example of a non-constant curvature Riemann surface on which one can compactify D and M branes, providing a holographic description of CFTs compactified on spindles. Moreover, this is a context were it is understood how to compute various CFT observables allowing one to make a direct comparison with gravity results. For examples of work in this vein see for instance \cite{Ferrero:2020laf,Ferrero:2021etw,Ferrero:2021wvk, Couzens:2021rlk,Arav:2022lzo, Couzens:2022yiv,Suh:2023xse,Amariti:2023mpg,Amariti:2023gcx, Bomans:2024mrf,Crisafio:2024fyc,Ferrero:2024vmz,Arav:2025jee,Suh:2025cii}. 

Another context where spindles have found an interesting application is for black holes. Famously asymptotically AdS$_4$ black holes evade the uniqueness theorem that forces asymptotically flat, non-rotating, black holes to contain a 2-sphere. As such it is natural to consider more general, possibly non-constant curvature, Riemann surfaces - a natural candidate is the spindle. Extremal black holes and extremal black hole near horizons containing spindles do indeed exist, and have been constructed and studied in \cite{Ferrero:2020twa,Ferrero:2021ovq,Boido:2022mbe,Couzens:2021cpk,Hristov:2023rel,Hristov:2024qiy,Colombo:2024mts,Boisvert:2024jrl,Hosseini:2021fge,Hosseini:2023ewi}.

Beyond supergravity, spindles have also been considered in the context of supersymmetric gauge theories defined on spaces that contain them and localisation, see for instance \cite{Inglese:2023wky,Inglese:2023tyc,Pittelli:2024ugf, Colombo:2024mts, Jeon:2025rfc}.\\

Given the above successes, people have naturally also considered related constructions. A closely related non-constant curvature Riemann surface is a topological disc which appeared prominently the holographic dual of Argyres-Douglas theories in \cite{Bah:2021mzw,Bah:2021hei}. It was later realised that one could generate such solutions from existing multi-charge spindle geometries \cite{Couzens:2022yjl} - see \cite{Couzens:2021tnv,Couzens:2021rlk,Suh:2021ifj,Suh:2021aik, Bomans:2023ouw} for other examples with topological discs. Along similar lines, multi charge spindle solutions have also been used as the starting point for the construction of holographic duals to superconformal monodromy defects in various CFTs \cite{Arav:2024exg,Arav:2024wyg,Conti:2025wwf}.

Another natural generalisation of the spindle is to consider similar orbifolds in higher dimensions. The most obvious thing one could aim for are solutions describing  branes wrapping $d=4$ orbifolds consisting of a product of two Riemann surfaces, at least one of which is a spindle. The next step up from that is to allow one of these Riemann surfaces to be fibred over the other, this has lead to works such as \cite{Couzens:2022lvg,Cheung:2022ilc,Suh:2022olh,Pittelli:2024ugf,Suh:2024fru}. More general $d=4$ orbifold can also be found in \cite{Faedo:2024upq}.

Another way to generalise the spindle, which will be the focus of this work, is to observe that like $\mathbb{CP}^1$, $\mathbb{WCP}^1$ is the first entry in a series of geometries. Just as $\mathbb{CP}^n$ exist for $n=1,2,...$, so too does $\mathbb{WCP}^n$ which is a space that is topologically $\mathbb{CP}^n$ with $n+1$ orbifold singularities parametrised by $n+1$ weights $(k_1,...k_{n+1})$. Indeed un-fibred\footnote{Specifically we refer to a manifold being fibred if it appears as the base of a fibre bundle containing a topologically non-trivial connection.} $\mathbb{WCP}^2_{[1,1,2]}$ and $\mathbb{WCP}^3_{[1,1,1,3]}$  have appeared before in a supergravity context in \cite{Gauntlett:2004zh,Bianchi:2021uhn}, while  $\mathbb{WCP}^2_{[k_1,k_2-k_1,k_2]}$ appeared in \cite{Macpherson:2024frt} as part of an effort to construct marginal deformations of the CFTs dual to Maldacena-Gaiotto geometries \cite{Gaiotto:2009gz} - each of these constructions preserve supersymmetry. Note that $\mathbb{WCP}^2$ has also appeared in a localisation context in \cite{Mauch:2024uyt}. Our aim in this work will be to construct examples of solution in $d=10$ and $d=11$ dimensional supergravity containing orbifolds involving $\mathbb{WCP}^2$ and $\mathbb{WCP}^3$ factors. Our method of achieving this is based on the fact $\mathbb{WCP}^n$ can be defined through a weighted circle action on S$^{2n+1}$ such that $\mathbb{WCP}^n=\text{S}^{2n+1}/\text{U}(1)$. Specifically one can embed the $(2n+1)-$sphere in $\mathbb{C}^{n+1}$ in terms of coordinates $z_a$ for $a=1,...n+1$ which are contained as $(z_a)^2=1$. From this one can define a weighted circle action on $\text{S}^{2n+1}$ as
\beq
(z_1,~....,~z_{n+1})\to (q^{k_1}z_1,~....,~q^{k_{n+1}}z_{n+1}),~~~~q=e^{i \theta}\in \text{U}(1)
\eeq
where $l\in\mathbb{Z}$ and $k_a\in \mathbb{N}$ - the quotient $\text{S}^{2n+1}/\text{U}(1)$ then gives rise to $\mathbb{WCP}^n_{[k_1,... k_{n+1}]}$. If we fix $k_1=...=k_{n+1}=1$ we recover $\mathbb{CP}^{n}$, but note that unlike this space, $\mathbb{WCP}^n$ is not Kahler-Einstein. From this weighted action one can then construct a, generically non free, $\mathbb{Z}_l$ action on the $\text{S}^{2n+1}$ which operates in the same manor as above but for $q= e^{\frac{2\pi i}{l}}$. Given this, it is possible to construct  non-trivial U(1) orbifold fibrations over $\mathbb{WCP}^n$ by performing some fairly simple operations on odd dimensional spheres, and to establish when such orbifolds preserve supersymmetry. Through this analysis it will become apparent that, unlike the spindle, both un-fibred $\mathbb{WCP}^2$ and $\mathbb{WCP}^3$ support supersymmetry for certain tunings of their weights, without a gauge field needing to enter the game.  

Starting from existing supergravity solutions containing odd dimensional spheres we will generate new supersymmetric AdS solutions containing U(1)-orbifold bundles over $\mathbb{WCP}^n$, and then through T-duality and IIA-M-theory duality, solutions containing un-fibred $\mathbb{WCP}^n$. This leads to interesting supersymmetric solutions such as AdS$_4\times \mathbb{WCP}^3$ and AdS$_5\times \mathbb{WCP}^2\times \text{S}^1$. We will also be able to construct a $d=5$ orbifold with 4 fixed points that generalises $\mathbb{T}^{(1,1)}$ in a similar fashion. Our focus will be on AdS solutions realisable as near horizon limits of certain brane systems and with well understood CFT duals. The reason for that is in part tractability on the gravity side, but we also suspect that these are the geometries containing $\mathbb{WCP}^n$ whose CFT duals are most likely to be understood in the future - an achievement we will not attempt here.\\
~\\
The lay out of this work is as follows:\\
~\\
In section \ref{sec:two} we derive U(1) orbifold bundles over $\mathbb{WCP}^n$ for $n=1,2,3$ starting from $(2n+1)$-spheres. We establish their supersymmetry properties and when they admit a supersymmetry preserving Kaluza-Klein reduction to un-fibred $\mathbb{WCP}^n$. For $\mathbb{WCP}^1$ we recover known results from the literature but for $\mathbb{WCP}^n$ with $n>1$, to our knowledge, our results are new. We will in fact construct two parametrisations of $\mathbb{WCP}^3$ based on the realisations of $\mathbb{CP}^3$ as a foliation over an interval of a squashed 5-sphere and $\mathbb{T}^{(1,1)}$ respectively.

In section \ref{sec: three} we construct non-trivial orbifolds of well known AdS near horizon geometries, study their flux quantisation and make the first step towards understanding their CFT duals by computing their, Wely anomalies, free energies or central charges, as appropriate to the dimensions at hand. We specifically construct orbifolds of the D3, D1-D5 and M5 and M2 near horizon geometries giving rise to AdS$_d$ solutions for $d=3,4,5,7$.

Section \ref{eq:solswithroundWCPn} utilises supergravity dualities to construct supersymmetric AdS solutions containing un-fibred $\mathbb{WCP}^n$ factors. Specifically we construct a supersymmetry breaking two parameter deformation of AdS$_7\times \text{S}^2\times I$ in section \ref{sec:M5branereduction}, supersymmetric AdS$_5\times\mathbb{WCP}^2\times \text{S}^1$ and AdS$_5\times\mathbb{WCP}^2\times \mathbb{T}^2$ in section \ref{sec:D3branereduction} and supersymmetric AdS$_4\times\mathbb{WCP}^3$ in section \ref{sec:AdS4wcp3}.

As the analysis is a bit different to the other cases, in section \ref{eq:WT11ex} we consider supersymmetric orbifolds of AdS$_3\times\text{S}^3\times\text{S}^3\times \text{S}^1$ containing an U(1)-orbifold bundle over a topological $\mathbb{T}^{(1,1)}$-like space we dub $\mathbb{WT}^{(1,1)}$. We establish that it is compatible with ${\cal N}=(2,2)$ supersymmetry generically and allows one to T-dualise to type IIA without breaking further supersymmetry. This results in a solution with an un-fibred $\mathbb{WT}^{(1,1)}$ factor.

Finally we make some concluding remarks and comment on some possible future directions in section \ref{sec:conclusions}.

\section{Supersymmertic $\mathbb{WCP}^n$ from $(2n+1)$-spheres} \label{sec:two}
In this section we construct orbifolded circle fibrations over the weighted projective spaces $\mathbb{WCP}^2_{[k_1,k_2,k_3]}$ and $\mathbb{WCP}^3_{[k_1,k_2,k_3,k_4]}$, which are higher dimensional analogues of the spindle, which is $\mathbb{WCP}^1_{[k_1,k_2]}$. After taking an orbifold of the fibre circle we establish under what conditions these U(1) orbifold bundles over $\mathbb{WCP}^n$ preserve supersymmetry and under what conditions it is possible to perform a Kaluza-Klein reduction on the U(1) and arrive at un-fibred $\mathbb{WCP}^n$ with some portion of supersymmetry intact. We study the case of $\mathbb{WCP}^2_{[k_1,k_2,k_3]}$ in section \ref{sec:WCP2} and two parametrisation of  $\mathbb{WCP}^3_{[k_1,k_2,k_3,k_3]}$ in sections \ref{sec:WCP31} and \ref{sec:WCP32}. Before that though, in section \ref{eq:wcp1sec},  we will review U(1) orbifold bundles over $\mathbb{WCP}^1$, as it will inform us going forward.

\subsection{$\mathbb{WCP}^1$ }\label{eq:wcp1sec}
The weighted projective space $\mathbb{WCP}^1_{[k_1,k_2]}$, where $k_i$ are integers we can take to be positive without loss of generality, is often referred to as a spindle. We can construct this, or rather a circle fibration over it, starting from the 3-sphere \cite{Ferrero:2020twa,Arav:2025jee}. We take the parametrisation
\beq
ds^2(\text{S}^3)=\frac{1}{4}d\theta^2+\sin^2\left(\frac{\theta}{2}\right) d\phi^2+\cos^2\left(\frac{\theta}{2}\right) d\psi^2,\label{eqn:S3}
\eeq
where the isometries $(\phi,\psi)$ have period $2\pi$. One can perform the following coordinate transformation of the isometry directions
\beq
\left(\begin{array}{c}\psi\\ \phi\end{array}\right)\to\left(\begin{array}{c c}k_1&m_1\\ k_2&m_2\end{array}\right)\left(\begin{array}{c}\psi\\ \phi\end{array}\right), \label{eq:S3coordtrans}
\eeq
which is an SL(2, $\mathbb{Z}$) transformation, and so preserves the periods of $(\phi,\psi)$, when  $k_1m_2-k_2m_1=1$ for $m_{1,2}\in\mathbb{Z}$. If we assume $\text{gcd}(k_1,k_2)=1$ then such an $(m_1,~m_2)$ do indeed exist via B\'ezout's identity. This leads to a circle fibration over a spindle with $\partial_{\psi}$ the fibre isometry, however given that all we have done is perform an SL$(2,\mathbb{Z})$ transformation of the U(1) isometry directions, the result at this point is nothing more than a reparametrisation of the 3-sphere. One can construct a non-trivial orbifolding of the 3-sphere by performing a $\mathbb{Z}_l$ orbifolding of the $\psi$ direction such that
\beq
\psi\to  \frac{1}{l}\psi, \label{eq:introl}
\eeq
for $l$ an integer. Let us stress that while we keep $l$ arbitrary during our analysis of the geometry in this and the following sections, not all values of $l$ here or there will be compatible with supersymmetry. The result of applying \eqref{eq:S3coordtrans} and \eqref{eq:introl} to the 3-sphere metric is to map it to the form
\begin{align}
ds^2(\mathbb{B}^3)=\frac{1}{4}\bigg(d\theta^2 +\frac{\sin^2\theta}{\Delta}d\phi^2\bigg)+\frac{\Delta}{l^2}(d\psi+\mathcal{A})^2,~~~~~~~~\Delta= k_1^2 \cos^2\left(\frac{\theta}{2}\right) +k_2^2\sin^2\left(\frac{\theta}{2}\right)\label{eq:orbibundelS3}
\end{align}
with the connection 1-form defined as 
\beq
\mathcal{A}=l \bigg( \frac{m_1}{k_1}+ \frac{k_2 \sin^2\left(\frac{\theta}{2}\right)}{k_1 \Delta}\bigg)d \phi,
\eeq
where $(\phi,\psi)$ still have periods $2\pi$ such that the space we have derived is a U(1) orbifold bundle over a spindle, which is the weighted projective space $\mathbb{WCP}^1_{[k_1,k_2]}$, where 
\beq
ds^2(\mathbb{WCP}^1)=\frac{1}{4}\bigg(d\theta^2 +\frac{\sin^2\theta}{\Delta}d\phi^2\bigg).\label{eq:WCP1def}
\eeq
Clearly $\mathbb{WCP}^1$ is topologically a 2-sphere with two $\mathbb{R}^2/Z_k$ orbifold singularities at its poles, \textit{i.e.}
\beq
\Delta(\theta=0)=k_1^2,~~~~\Delta(\theta=\pi)=k_2^2\label{eq:deltabehavour},
\eeq
with $\Delta$ continuous and neither blowing up nor vanishing between these bounds. It is straight forward to compute the volume of $ds^2(\mathbb{WCP}^1)$, we find
\beq
\int_{\mathbb{WCP}^1}\text{vol}(\mathbb{WCP}^1)= \frac{2\pi}{k_1+k_2},\label{eq:volWCP1}
\eeq  
while integrating the field strength of the connection 1-form over this space leads to 
\beq
\frac{1}{2\pi}\int_{\mathbb{WCP}^1}d\mathcal{A}=\frac{l}{k_1k_2},\label{eq:char2d}
\eeq
The Euler characteristic on the spindle may be easily calculated through the identity
\beq
\chi_E=\frac{1}{4\pi} \int_{\mathbb{WCP}^1 }R\,\text{vol}(\mathbb{W}\mathbb{C}\mathbb{P}^1)=\frac{k_1+k_2}{k_1k_2}=2-\bigg(1-\frac{1}{k_1}\bigg)-\bigg(1-\frac{1}{k_2}\bigg),\label{eq:char2dr}
\eeq
for $R$ the Ricci scalar of $ds^2(\mathbb{WCP}^1)$, which yield the expected rational results.

Of course $\mathbb{WCP}^1$ is a topological space, there can be, and do exist, different manifolds with this topology. Indeed starting from the squashed 3-sphere
\beq
ds^2(\text{S}^3)=\frac{1}{4}(d\theta^2+\sin^2\theta d\phi^2)+\lambda^2(d\psi+\sin^2\left(\frac{\theta}{2}\right) (d\phi-d\psi))^2
\eeq
for $\lambda$ some constant  and then again operating with  \eqref{eq:S3coordtrans} and orbifolding as in \eqref{eq:introl}   one is mapped to 
\begin{align}
ds^2(\mathbb{B}^3_{\lambda})&=\frac{1}{4}\bigg(d\theta^2 +\frac{\lambda^2 \sin^2\theta}{\Delta_{\lambda}}d\phi^2\bigg)+\frac{\lambda^2\Delta_{\lambda}}{l^2}(d\psi+\mathcal{A}_{\lambda})^2, ~~~\Delta_{\lambda}=\Delta+\frac{1}{4\lambda^2}(1-\lambda^2)(k_1^2-k_2)^2\sin^2\theta,\nn\\[2mm]
\Delta_{\lambda}{\cal A}_{\lambda}&=\Delta{\cal A}-l(1-\lambda^2)\left(k_1 \sin^2\left(\frac{\theta}{2}\right)+k_2 \cos^2\left(\frac{\theta}{2}\right)\right)\left(m_1 \sin^2\left(\frac{\theta}{2}\right)+m_2 \cos^2\left(\frac{\theta}{2}\right)\right) d\phi.\label{eq:orbibundelS3def}
\end{align}
Clearly this is a parametric deformation of a $\mathbb{B}^3$ in terms of $\lambda$, but $(\lambda^{-2}\Delta_{\lambda},~{\cal A}_{\lambda})$ behave the same as $(\Delta,{\cal A})$ at the poles of the $(\theta,\phi)$ spanning deformed S$^2$, and neither of $(\Delta_{\lambda},~{\cal A}_{\lambda})$ can blow up or shrink to zero for real $\theta$.  We can thus equally well identify the base of \eqref{eq:orbibundelS3def} as $\mathbb{WCP}^1$. Indeed, inserting the above base into \eqref{eq:char2d}  also leads to \eqref{eq:char2dr} - however the volume and Ricci scalar of this version  of $\mathbb{WCP}^1$ depend on $\lambda$, so it is a distinct space. We stress this point because we will construct higher dimensional weighted projective spaces and study their supersymmetry properties in the following sections. As will become apparent, these spaces are foliations that contain ${\mathbb B}^3$ factors -- \textit{i.e} $\mathbb{WCP}^2$ can be viewed as a foliation of a squashed ${\mathbb B}^3$ over an interval with ${\mathbb B}^3$ also appearing in the construction of $\mathbb{WCP}^3$. But it should be possible to construct distinct versions of $\mathbb{WCP}^n$ that use distinct manifolds with the same topology as ${\mathbb B}^3$ as their building blocks.

In the next subsection we will study the supersymmetry properties of $\mathbb{B}^3$ which will inform our later analyses for $\mathbb{WCP}^{2,3}$ .

\subsubsection{Supersymmetry}\label{eq:WCP1susy}
Now to address the issue of supersymmetry preservation we can consider the Killing spinors on the 3-sphere. There are two sets of Killing spinors on S$^3$: $\xi^{(3)}_{\pm}$ that solve the Killing spinor equations
\beq
\nabla_{m}\xi^{(3)}_{\pm}=\pm\frac{i}{2}\gamma^{(3)}_{m}\xi^{(3)}_{\pm}.
\eeq
With respect to the following frame and $\gamma$-matrices 
\beq
e^{a}=\left(\frac{1}{2}d\theta,~\frac{1}{2}\sin\theta (d\phi-d\psi),~d\psi+\sin^2\left(\frac{\theta}{2}\right) (d\phi-d\psi)\right)^{a},~~~~ \gamma^{(3)}_a=\sigma_a,\label{eq:nonnaturalframeB3}
\eeq
these take the form
\beq
\xi^{(3)}_{\pm}= {\cal M}^{(3)}_{\pm}\eta^{(3)}_{\pm},~~~{\cal M}^{(3)}_+= e^{\frac{i}{2}(\phi+\psi) \gamma_3},~~~~{\cal M}^{(3)}_-= e^{-\frac{i}{2} \theta \gamma_1}e^{\frac{i}{2}(\phi-\psi)\gamma_{12}}, \label{eq:S3spinors}
\eeq
where $\eta^{(3)}_{\pm}$ are arbitrary constant spinors. After performing the transformation \eqref{eq:S3coordtrans} a natural frame on  ${\mathbb B}^3$ is
\begin{equation}
\hat e^{a} = \left( \frac{1}{2} d \theta,~ \frac{1}{2} \frac{\sin \theta}{\sqrt{\Delta}} d \phi,~ \frac{\sqrt{\Delta}}{l}(d \psi + {\cal{A}})\right)^a,\label{eq:B3viel}
\end{equation}
which is related to \eqref{eq:nonnaturalframeB3} by first applying \eqref{eq:S3coordtrans} and \eqref{eq:introl} and then through the SO(2) transformation
\begin{equation}
\hat e^{a} = {\cal R}^{a}_{~b} \, e^{b}, \qquad {\cal R} = 
\left(\begin{array}{ccc}1&0&0 \\
0 & \cos X & -\sin X \\
0 & \sin X & \cos X
\end{array}\right),
\end{equation}
where the relevant angle of rotation is
\begin{equation}
\tan X = -\frac{(k_1 - k_2) \sin \theta}{k_1 + k_2 + (k_1 - k_2)\cos \theta}. \label{eq:Xdef}
\end{equation}
The spinors in the frame \eqref{eq:B3viel} are defined in terms of the spinors on S$^3$ as 
\begin{equation}
\hat{\xi}^{(3)}_{\pm} = \Omega \, \xi^{(3)}_{\pm}
\end{equation}
where $\Omega$ obeys and is equal to
\begin{equation}
\Omega^{-1} \gamma^{a} \Omega = {\cal R}^{a}_{~b} \, \gamma^{a},~~~~\Omega = e^{- \frac{X}{2} \gamma_{23} },
\end{equation}
and \eqref{eq:S3coordtrans} and \eqref{eq:introl} must be applied. Explicitly we find that the spinors in the natural frame for $\mathbb{B}^3$ take the form
\begin{align}
\hat{\xi}^{(3)}_{\pm} &= \hat{\cal M}^{(3)}_{\pm} \eta^{(3)}_{\pm}, \qquad\hat{\cal M}^{(3)}_{+}=e^{- \frac{X}{2} \gamma_{23} }e^{\frac{i}{2}\left((m_1+m_2)\phi+\frac{(k_1+k_2)}{l}\psi\right) \gamma_{3}},\nn\\[2mm]
\hat{\cal M}^{(3)}_{-}  &= e^{- \frac{X}{2} \gamma_{23} }e^{-\frac{i}{2}\theta\gamma_1}e^{-\frac{1}{2}\left((m_1-m_2)\phi+\frac{(k_1-k_2)}{l}\psi\right) \gamma_{12}}.\label{eq:spinorsB3}  
\end{align}
For supersymmetry to be preserved globally we need these spinors to be well defined globally. First off we should have that under $\psi\to \psi +2\pi$ the spinor should be mapped to either itself or minus itself  which, given that $\text{gcd}(k_1,k_2)=1$, this can only happen for both of $\hat{\xi}^{(3)}_{\pm}$ if $l=1$.  However  either $\hat{\xi}^{(3)}_{+}$ or $\hat{\xi}^{(3)}_{-}$ alone will also satisfy this property if we fix $l=k_1\pm k_2$, up to an overall sign one can absorb by sending $\psi\to 2\pi-\psi$. One can also fix $k_1=k_2=1$ and ensure this for $\hat{\xi}^{(3)}_{-}$, but this results in S$^3/\mathbb{Z}_l$ which is not of interest to us here. Thus supersymmetry requires that either
\beq
l=1,~~~~\text{or}~~~~l=k_1\pm k_2 \label{eq:lrule1}.
\eeq
Note however that for $l=1$ we have not really done anything other than rewrite S$^3$ in an overly complicated fashion. The other choices give rise to the twist and anti-twist realisations of supersymmetry on a spindle  discussed in \cite{Ferrero:2021etw}. 
Since $(k_1,k_2,m_1,m_2)$ are all integers we manifestly have that $\hat{\xi}^{(3)}_{\pm}$ behaves properly as $\phi\to \phi+2\pi$, but at the poles of $\mathbb{WCP}^1$, where we have orbifold singularities, this coordinate is not defined and so things are a bit more subtle. Specifically, as explained in \cite{Ferrero:2021etw}, we need to ensure that the spinors are regular in the covering spaces for the orbifolds at the poles, which take the form $\mathbb{R}^2\times$S$^1$. Close to the respective poles $\mathbb{B}^3$ approaches
\beq
ds^2= \frac{1}{4}(dr^2+ r^2 d\hat \phi_i^2)+ \frac{k_i^2}{l^2}(d\psi+ l m_i d\hat{\phi}_i)^2,~~~~ \hat{\phi}_i=\frac{1}{k_i}\phi,   
\eeq 
where $r=\theta$ as $\theta\to 0$ and $r=\pi-\theta$ as $\theta\to \pi$. On $\mathbb{B}^3$ the gauge field becomes ${\cal A}= l m_i  d\hat\phi_i=l\frac{m_i  }{k_i}d\phi$ at the poles which is singular, but on the covering space $\hat\phi_i$ have period $2\pi$ and there one can define a new coordinate $\chi_i= \psi+ l m_i \hat\phi_i$ through an SL(2, $\mathbb{Z}$) transformation, arriving at
\beq
ds^2= \frac{1}{4}(dr^2+ r^2 d\hat \phi_i^2)+ \frac{k_i^2}{l^2}d\chi_i^2,
\eeq
which is regular on the covering space where $(\hat \phi_i,\chi_i)$ both have period $2\pi$. In terms of these coordinates we find that the spinors on $\mathbb{B}^3$ in the frame of \eqref{eq:B3viel} tend at the poles to
\begin{align}
\theta\sim 0&:~~~\hat \xi_+^{(3)}=e^{\frac{1}{2}\left(\hat\phi_1+\frac{(k_1+k_2)}{l} \chi_1\right)\gamma_{12}}\eta^{(3)}_{+},~~~~\xi_-^{(3)}=e^{\frac{1}{2}\left(\hat\phi_1-\frac{(k_1-k_2)}{l} \chi_1\right)\gamma_{12}}\eta^{(3)}_{-},\nn\\[2mm]
\theta\sim \pi&:~~~\hat \xi_+^{(3)}=e^{\frac{1}{2}\left(-\hat\phi_2+\frac{(k_1+k_2)}{l} \chi_2\right)\gamma_{12}}\eta^{(3)}_{+},~~~~\xi_-^{(3)}=e^{\frac{1}{2}\left(-\hat\phi_2+\frac{(k_1-k_2)}{l} \chi_2\right)\gamma_{12}}e^{-\frac{\pi}{2}\gamma_{23}}\eta^{(3)}_{-}.
\end{align}
With respect to the frame \eqref{eq:B3viel}, evaluated at the poles, the $\hat \phi_i$ dependence in the above is precisely that of a spinor at the origin or $\mathbb{R}^2$ expressed in polar coordinates. Thus the spinors are indeed regular on the covering space proved that $l$ is tuned such that the $\psi$ coordinate is well behaved as previously discussed. Thus we see that supersymmetry is preserved globally on $\mathbb{B}^3$ when \eqref{eq:lrule1} is imposed.

Note that we necessarily have that  $\hat\xi^{(3)}_{\pm}$ are charged under $\partial_{\psi}$ unless $k_1=k_2=1$, this means that we cannot Kaluza-Klein reduce on $\partial_{\psi}$ within ${\mathbb B}^3$ and arrive at an un-fibred spindle geometry that preserves supersymmetry, like you can for  S$^3/\mathbb{Z}_l\to $S$^2$. This should be contrasted with the higher dimensional weighted projective spaces in the following sections which do allow this. One can of course Scherk-Schwarz reduce $\mathbb{B}^3$ on $\partial_{\psi}$ which maintains the gauge field ${\cal A}$ in the conditions for supersymmetry and so allows un-fibred $\mathbb{WCP}^1$ to be embedded in \textbf{gauged} supergravity.

As a final comment let us point out that our above analysis holds equally well for \eqref{eq:orbibundelS3def}. Indeed if we replace $e^3$ with $\lambda e^3$ in \eqref{eq:nonnaturalframeB3} they give a vielbein on the squashed S$^3$ this is constructed from. The spinors on this space are actually identical to \eqref{eq:S3spinors}, it is the equations they obey that get modified to 
\beq
\nabla_{m}\xi^{(3)}_+=(2-\lambda)\frac{i}{2}\gamma_m\xi_++i (\lambda-1)e^3_{m}\sigma_3\xi^{(3)}_+,~~~\nabla_{m}\xi^{(3)}_-=\lambda\frac{i}{2}\gamma_m\xi_-+i \frac{(\lambda^2-1)}{\lambda}e^3_{m}\sigma_3\xi^{(3)}_+.\label{eq:squshedS3eqs} 
\eeq 
After operating with  \eqref{eq:S3coordtrans} a natural frame on $\mathbb{B}_{\lambda}^3$ is $\hat e^a=(d\theta,~\frac{\sin\theta}{\Delta_{\lambda}}d\phi,~\frac{\lambda}{l}(d\psi+{\cal A}_{\lambda}))^a$ and the spinors take the form of \eqref{eq:spinorsB3} with one modification - $X$ as defined in \eqref{eq:Xdef} now gets its RHS multiplied by $\lambda^{-1}$. The rest of the discussion involving the preservation of global supersymmetry then goes through the same.

\subsection{$\mathbb{WCP}^2$ }\label{sec:WCP2}
In this section we derive a metric on $\mathbb{WCP}^2_{[k_1,k_2,k_3]}$ which is a 4 dimensional space that is topologically $\mathbb{CP}^2$ with 3 orbifold singularities related to the positive integers $k_i$ for $i=1,2,3$. We will again actually construct an orbifolded circle fibration over this space, however we shall see that, unlike its lower dimensional cousin and for that matter $\mathbb{CP}^2$, it is possible to reduce on this circle without breaking all of the supersymmetry of the fibred space when $k_i$ are appropriately tuned. What this means is that certain un-fibred $\mathbb{WCP}^2_{[k_1,k_2,k_3]}$ manifolds support a spin structure, even though $\mathbb{WCP}^2_{[1,1,1]}=\mathbb{CP}^2$ does not.\\
~\\
We begin our derivation from the metric on the round 5-sphere expressed as a foliation of S$^1\times$S$^3$ over an interval, its metric takes the form
\begin{align}
ds^2(\text{S}^5)= d\mu^2+ \cos^2\mu \, d\beta^2+ \sin^2\mu \,ds^2(\text{S}^3).\label{eq:S5}
\end{align}
with the metric on S$^3$ expressed as in \eqref{eqn:S3}, giving us 3 U(1) isometries to work with. We operate on these via the SL$(3, \mathbb{Z}$) transformation
\beq
\left(\begin{array}{c}\psi\\ \phi\\\beta\end{array}\right)\to  \left(\begin{array}{ccc}0&m_1&k_1\\0&m_2&k_2\\-1&0&k_3\end{array}\right)\left(\begin{array}{c}\psi\\ \phi\\\beta\end{array}\right), \label{eq:WCP2step1}
\eeq  
where that $k_1m_2-k_2m_1=1$. We then orbifold the $\partial_{\beta}$ circle as
\beq
\beta\to \frac{1}{l}\beta. \label{eq:WCP2step2}
\eeq
This maps S$^5 \to \mathbb{B}^5$ with metric of the form
\begin{equation}
ds^2 (\mathbb{B}^5) = d\mu^2 + \frac{1}{4} \sin ^2\mu\left(d\theta^2 +  \frac{\sin ^2\theta }{\Delta} d\phi^2\right) + \frac{ \Delta \sin ^2\mu \cos ^2\mu }{\Xi} (d\psi+{\cal A})^2  +  \frac{\Xi}{l^2} \left(d \beta + {\cal{B}} \right)^2,\label{eq:defB5}
\end{equation}
where $(\beta,\psi,\phi)$ have period $2\pi$, we define the nowhere vanishing functions
\begin{align}
\Delta = k_1^2 \cos ^2\left(\frac{\theta }{2}\right) + k_2^2 \sin ^2\left(\frac{\theta }{2}\right), \qquad \Xi = \Delta \sin ^2 \mu + k_3^2 \cos ^2 \mu,
\end{align}
and the connections 1-forms  are given by
\begin{equation} \label{eq:defconnections}
{\cal A}=  k_3\bigg( \frac{m_1}{k_1}+ \frac{k_2 \sin^2\left(\frac{\theta}{2}\right)}{k_1 \Delta}\bigg)d \phi ,~~~~ {\cal B}= \frac{l}{k_3} \frac{\Delta}{\Xi } \sin ^2 \mu ( d \psi + {\cal A})-\frac{l}{k_3} d\psi,
\end{equation}
note that ${\cal A}$ takes the same form as it did on $\mathbb{B}^3$ but with $l=k_3$. This space is a U(1) orbifold bundle over a base manifold that is topologically $\mathbb{CP}^2$ with 3 orbifold singularities with associated integers $(k_1,k_2,k_3)$, \textit{i.e.} the base manifold is the weighted projective space $\mathbb{WCP}^2_{[k_1,k_2,k_3]}$, we thus identify
\begin{align}
ds^2(\mathbb{WCP}^2)&=d\mu^2 + \frac{1}{4} \sin ^2\mu ds^2(\mathbb{WCP}^1) + \frac{ \Delta \sin ^2\mu \cos ^2\mu }{\Xi} D\psi^2,\nn\\[2mm]
ds^2(\mathbb{WCP}^1)&= d\theta^2 +  \frac{\sin ^2\theta }{\Delta} d\phi^2,~~~D\psi=d\psi+{\cal A}.\label{eq:WCP2def}
\end{align}
This identification can be justified as follows: First we observe that the above metric reproduces the metric on $\mathbb{CP}^2$ when $k_1=k_2=k_3=1$, \textit{i.e.} 
\beq
ds^2(\mathbb{CP}^2)= d\mu^2+\sin^2\mu \, ds^2(\mathbb{CP}^1)+ \sin^2\mu\cos^2\mu(d\psi+ \eta)^2,~~~~d\eta=-2\text{vol}(\mathbb{CP}^1).
\eeq
Notice that at $\mu=0$ this vanishes as the origin of $\mathbb{R}^4$ expressed as a regular cone over S$^3$, while at $\mu=\frac{\pi}{2}$ we have that $\mathbb{CP}^1$ has constant radius while the fibre shrinks smoothly. Returning to \eqref{eq:WCP2def} we note that as $\Delta$ clearly still behaves as in \eqref{eq:deltabehavour},  the $(\mathbb{R}^2/\mathbb{Z}_{k_1},\mathbb{R}^2/\mathbb{Z}_{k_2})$ orbifold singularities of $\mathbb{WCP}^1$ are lifted to $\mathbb{WCP}^2$. The third orbifold singularity becomes evident from the behaviour of $(\Xi,~{\cal B})$ at the boundaries of $\mu$, specifically
\beq
(\Xi,~{\cal B})\bigg\lvert_{\mu=0}=(k_3^2,-\frac{l}{k_3}d\psi),~~~~~(\Xi,~{\cal B})\bigg\lvert_{\mu=\frac{\pi}{2}}= (\Delta,l{\cal A}).
\eeq
This makes the space smooth at $\mu=\frac{\pi}{2}$ like it is for $\mathbb{CP}^2$, but at $\mu=0$ it is vanishing like the origin of a cone of base $\mathbb{B}^3$ as defined in \eqref{eq:orbibundelS3} with $l=k_3$, ${\cal B}$ is also singular here like ${\cal A}$ was on $\mathbb{B}^3$. Thus we conclude that the identification in \eqref{eq:WCP2def} is correct.

We find that the volume of the two manifolds are given by
\begin{align}
\int \text{vol}(\mathbb{B}^5) & = \frac{\pi ^3}{l }, \\[2mm]
\int \text{vol}(\mathbb{WCP}^2) & = \frac{4\pi^2}{3}\frac{k_1 k_2 + k_1k_3 + k_2 k_3}{ (k_1 + k_2 ) (k_1 + k_3) (k_2 + k_3)},
\end{align}
where the first of these is merely the same volume one would get on S$^5/\mathbb{Z}_l$. Integrating the field strengths of the two connections over $\mathbb{WCP}^1_{[k_1,k_2]}$, gives
\begin{equation}
\frac{1}{2\pi}\int_{\mathbb{W}\mathbb{C}\mathbb{P}^1}d\mathcal{A}=  \frac{k_3}{k_1k_2}, \qquad \frac{1}{2\pi}\int_{\mathbb{W}\mathbb{C}\mathbb{P},\mu=\frac{\pi}{2}} d\mathcal{B}=  \frac{l}{k_1k_2}.
\end{equation}
Finally to further confirm that the base manifold really is $\mathbb{WCP}^2_{[k_1,k_2,k_3]}$ we can compute the Euler characteristic which is given by
\begin{align}
\chi_E&=\frac{1}{32\pi^2} \int_{\mathbb{W}\mathbb{C}\mathbb{P}^2} (R_{abcd}R^{abcd}-4R_{ab}R^{ab}+R^2)\text{vol}(\mathbb{W}\mathbb{C}\mathbb{P}^2),
\end{align}
where the curvature tensors are computed with respect to $ds^2(\mathbb{WCP}^2)$. After a lengthy but rather pedestrian computation we find that
\begin{equation}
\chi_E=\frac{k_1 k_2+k_1 k_3+ k_2 k_3}{k_1 k_2 k_3 }=3-\bigg(1-\frac{1}{k_3}\bigg)-\bigg(1-\frac{1}{k_1}\bigg)-\bigg(1-\frac{1}{k_2}\bigg),
\end{equation}
consistent with our claim.

\subsubsection{Supersymmetry}\label{eq:wcp2susy}
We would now like to establish under what conditions supersymmetry is preserved by $\mathbb{B}^5$. Contrary to the spindle we will also show that it is possible to dimensionally reduce $\mathbb{B}^5$ to $\mathbb{WCP}^2$ without breaking supersymmetry entirely.\\
~\\
As it is locally the 5-sphere, clearly $\mathbb{B}^5$ should locally support the same Killing spinors as S$^5$, namely those obeying the equation\footnote{Note that we do not need to also consider a minus sign in this equation like we did for the 3-sphere, because in this case the basis of spinors solving  $\nabla_{\mu}\xi^{(5)}_-= -\frac{i}{2}\gamma_{\mu}\xi^{(5)}_-$ is simply the Majorana conjugate of the basis solving  $\nabla_{\mu}\xi^{(5)}_+= +\frac{i}{2}\gamma_{\mu}\xi^{(5)}_+$. }
\beq
\nabla_{m}\xi^{(5)}=\frac{i}{2}\gamma_{m}\xi^{(5)}.\label{eq:S5Killingspinoreq}
\eeq
We begin by taking the following vielbein and gamma matrix basis for the round 5-sphere
\begin{align}
e^{a}&=\left(d\mu,~\cos\mu d\beta,~\frac{1}{2} \sin\mu d\theta,~  \sin\mu \sin\theta (d\phi-d\psi),~ \sin\mu \left(d\psi+ \sin^2\left(\frac{\theta}{2}\right) (d\phi-d\psi) \right)\right)^{a},\nn\\[2mm]
\gamma_1&= \sigma_1\otimes \mathbb{I},~\gamma_2= \sigma_2\otimes \mathbb{I},~~~\gamma_{i+3}=\sigma_3\otimes \sigma_i,~~~i=1,2,3.\label{eq:S5viel}
\end{align}
With respect to this, it is a straightforward exercise to confirm that the Killing spinor equation is solved by
\begin{align}
\xi^{(5)}&=e^{\frac{i}{2}\mu \gamma_1}\bigg(e^{-\frac{i}{2}\beta}(\mathbb{I}\otimes {\cal M}^{(3)}_{+}) P_{-}+e^{\frac{i}{2}\beta}(\mathbb{I}\otimes {\cal M}^{(3)}_{-}) P_{+}\bigg)\eta^{(5)}, ~~~~ P_{\pm}=\frac{1}{2}(\mathbb{I}\pm \gamma_2),\label{eq:S5spinor}
\end{align}
where $\eta^{(5)}$ is an arbitrary constant spinor in 5 dimensions and ${\cal M}^{(3)}_{\pm}$ are the matrices that appear in the 3-sphere Killing spinors in \eqref{eq:S3spinors}. We can generate the Killing spinors on $\mathbb{B}^5$ by performing the sequence of operations in  \eqref{eq:WCP2step1}-\eqref{eq:WCP2step2}. We elect the following frame on $\mathbb{B}^5$
\beq
\hat{e}^{a}=\left(d\mu,~\frac{1}{2}\sin\mu d\theta,~\frac{1}{2}\sin\mu \frac{\sin\theta}{\sqrt{\Delta}}d\phi,~\frac{\Delta}{\Xi}\sin\mu\cos\mu D\psi,~\frac{\sqrt{\Xi}}{l}D\beta\right)^{a},\label{eq:B5frame}
\eeq
which is related to the vielbein in \eqref{eq:S5viel} by applying \eqref{eq:WCP2step1}-\eqref{eq:WCP2step2}, performing a proper Lorentz transformation of the frame and then sending $(\psi,k_3)\to-(\psi,k_3)$, which has no effect on the metric. In this frame the spinors on $\mathbb{B}^5$ take the form\footnote{Strictly speaking the $\eta^{(5)}$ appearing below is a constant rotation of the $\eta^{(5)}$ appearing in \eqref{eq:S5spinor}.}
\begin{align}
\hat \xi^{(5)}&=e^{-\frac{Y}{2}\gamma_{45}} e^{\frac{i}{2}\mu\gamma_1} e^{-\frac{1}{2} X \gamma_{35}}\bigg(e^{-\frac{i}{2l}(k_3\beta-l \psi)}{\cal N}_{+}\hat P_{+}+e^{\frac{i}{2l}(k_3\beta-l \psi)}{\cal N}_{-}\hat P_{-}\bigg)\eta^{(5)},~~~~~\hat{P}_{\pm}=\frac{1}{2}(\mathbb{I}\pm \gamma_4)\nn\\[2mm]
{\cal N}_{+}&=e^{\frac{1}{2l}((k_1+k_2)\beta+l(m_1+m_2)\phi)\gamma_{23}},~~~~{\cal N}_{-}=e^{-\frac{1}{2}\theta\gamma_{35}}e^{-\frac{1}{2 l }((k_1-k_2)\beta+l(m_1-m_2)\phi)\gamma_{23}}\label{eq:B5spinors}
\end{align}
where the two functions $(X,Y)$ are defined as
\beq
\tan X=-\frac{(k_1-k_2)\sin\theta}{k_1+k_2+(k_1-k_2)\cos\theta},~~~~\tan Y=-\frac{k_3 \cot\mu}{\sqrt{\Delta}}.\label{eq:spinspin}
\eeq
By construction $\hat \xi^{(5)}$ contain 4 independent spinors solving \eqref{eq:S5Killingspinoreq}, to establish whether they are well defined globally we find it useful to decompose $\eta^{(5)}$ in terms of eigenspinors of both $\gamma_4$ and  $i \gamma_{23}$ as
\beq
\eta^{(5)}=\eta^{(5)}_{++}+\eta^{(5)}_{+-}+\eta^{(5)}_{-+}+\eta^{(5)}_{--}
\eeq
where the notation $\eta^{(5)}_{pq}$ means that $p$ is the eigenvalue under $\gamma_4$ and $q$ under $i \gamma_{34}$. In this way we isolate 4 independent spinors on $\mathbb{B}^5$, $\xi^{(5)}_{pq}$ for $p,q=\pm$. Each of these spinors depends on $(\phi,\psi,\beta)$ only in terms of overall phases we can collect in terms of functions of the form $e^{i \Theta_{pq}}$, we have specifically that
\begin{align}
\Theta_{+\pm}&=\pm\frac{1}{2 l}\left((k_1-k_2\mp k_3)\beta+l(m_1-m_2)\phi\pm l\psi\right),\nn\\[2mm]
\Theta_{-\pm}&=\mp\frac{1}{2 l}\left((k_1+k_2\mp k_3)\beta+l(m_1+m_2)\phi\pm l\psi\right).
\end{align}
Again we should have that the spinor is well behaved as we traverse these U(1) circles which requires the coefficients of the $(\beta,\psi,\phi)$ terms in $\Theta_{pq}$ to be in $ \frac{1}{2}\mathbb{Z}$. The issue is the $\beta$ dependence, namely for generic values of $k_i$ this only happens for all 4 independent spinors if $l=1$, yielding the round 5-sphere. We can also make a single spinor well defined by fixing $l$ to one of 
\beq
l= (k_1-k_2+ k_3),~l= (k_1-k_2- k_3),~ l=(k_1+k_2- k_3),~ l=(k_1+k_2+ k_3),\label{eq:WCP2ls}
\eeq
which generically preserves $\frac{1}{4}$ of the supercharges on the 5-sphere and are the analogues of twist and anti-twist on the spindle. The form of $\Theta_{pq}$ makes it clear that it is possible to make choices of $k_i$ for which one of $\Theta_{pq}$ becomes independent of $\beta$, under the assumption that $k_i$ are all positive these are
\beq
k_3=k_1+k_2,~~~~k_1=k_2+k_3,~~~~ k_2=k_1+k_3,\label{eq:WCP2kitunings}
\eeq
each of which makes the $\beta$ dependence drop out of 1 of the 4 -spinors, allowing us to preserve $\frac{1}{4}$ of the supercharges for generic values of $l$. Given one of the choices in \eqref{eq:WCP2kitunings} it is also possible preserve $\frac{1}{2}$ of the supercharges for certain tunings of  $l$. For instance if we fix the weights as $k_3=k_1+k_2$ then the coefficient of the $\beta$ term in $\Theta_{pq}$ is given by
\beq
k_3=k_1+k_2:~~~\partial_{\beta}\theta_{++}=-\frac{k_2}{l},~~~\partial_{\beta}\Theta_{+-}=-\frac{k_1}{l},~~~\partial_{\beta}\Theta_{-+}=0,~~\partial_{\beta}\Theta_{--}=\frac{k_1+k_2}{l}.
\eeq
So for each of $l\in \{k_1,~k_2,~k_1+k_2\}$, a second phase shares the periodicity of $e^{i \Theta_{-+}}$ under $\beta\to \beta+2\pi$ yielding two well behaved spinors under this action. Beyond this, an interesting thing about the tunings in \eqref{eq:WCP2kitunings} is they imply that, for any value of $l$, one can Kaluza-Klein reduce on $\partial_{\beta}$ and preserve a single supercharge on un-fibred
\beq
\mathbb{WCP}^2_{[k_1,k_2,k_1+k_2]},~~~~\mathbb{WCP}^2_{[k_1,k_1+k_2,k_2]},~~~~~\mathbb{WCP}^2_{[k_1+k_2,k_1,k_2]},\label{eq:reductiontunign}
\eeq
we will make use of this fact in  section \eqref{sec:D3branereduction}.

We still need  to confirm that $\hat \xi^{(5)}$ is well behaved at the boundaries of the $\theta$ and $\mu$ intervals where respectively $\phi$ and $\psi$ are not defined. Close to the poles of the $\mathbb{WCP}^1$ factor, $\mathbb{B}^5$ approaches
\begin{align}
ds^2&= d\mu^2+ \cos^2\mu (d\chi_i-\frac{k_3}{l}d\tilde{\beta_i})^2+ \sin^2\mu\bigg[ \frac{1}{4}(dr^2+r^2d\hat \phi_i^2)+\frac{k_i^2}{l^2} d\tilde{\beta}_i^2\bigg],\nn\\[2mm]
\phi_i&=\frac{1}{k_i} \phi,~~~\chi_i=\psi+k_3 m_2\phi_i,~~~\tilde{\beta}_i=\beta+ l m_i \phi_i,
\end{align}
where the relevant term is the one in square brackets. On the $\mathbb{R}^2\times$S$^1$ covering spaces for the $\mathbb{R}^2/\mathbb{Z}_{k_i}$ orbifolds, $\hat\phi_i$ has period $2\pi$ and $(\chi_i,~\tilde{\beta}_i)$ can be defined by a sequence of SL$(2,\mathbb{Z})$ transformations. We see that the metric is regular on the covering space\footnote{At least when neither $\cos\mu$ or $\sin\mu$ are vanishing, limits we will consider momentarily},  and given that $Y$ becomes independent of $\theta$ at the poles, $k_3\beta-l \psi$ is independent of $\hat\phi_i$ and $e^{-\frac{1}{2} X \gamma_{35}}{\cal N}_{\pm}$ has essentially  the same structure as $\hat{\cal M}^{(3)}_{\pm}$ in \eqref{eq:spinorsB3}, it is not hard to establish that $\xi^{(5)}$ is also regular at the poles of the spindle for the same reason $\xi^{(3)}_{\pm}$ where.  Close to $\mu=\frac{\pi}{2}$,  $\mathbb{ B}^5$ tends to
\beq
ds^2= ds^2(\mathbb{WCP}^1_{[k_1,k_2]})+ \frac{\Delta}{l}(d\beta+\frac{l}{k_3}{\cal A})^2+  dr^2+r^2(d\psi+k_3 {\cal A})^2,~~~r=\frac{\pi}{2}-\mu,
\eeq
which is perfectly regular behaviour and so  the spinor is well defined here as the $\psi$ dependence in $\Theta_{pq}$ is in $\frac{1}{2}\mathbb{Z}$. The final orbifold singularity is at $\mu=0$ were the metric on $\mathbb{B}^5$ becomes
\beq
ds^2=dr^2+ r^2\bigg[ ds^2(\mathbb{WCP}^1_{[k_1,k_2]})+\Delta\left(d\hat \psi+ {\cal A}\right)^2\bigg]+ \frac{1}{l^2}(d\beta-d\hat\psi)^2,~~~~\hat\psi=\psi+\frac{l}{k_3}\phi.
\eeq
This is a $\mathbb{R}^4/\mathbb{Z}_k$ orbifold, so the covering space at this point is $\mathbb{R}^4\times$S$^1$ where $\hat\phi$ has period $2\pi$ and $\hat\beta$ can be defined via an SL$(2,\mathbb{Z})$ transformation. One can see that the solution is in fact regular on the covering space at this point in terms of the coordinates
\beq
x_1+i x_2= \cos\left(\frac{\theta}{2}\right)e^{-i(m_1\phi+k_1 \hat\psi)},~~~x_3+i x_4= \sin\left(\frac{\theta}{2}\right)e^{-i(m_2\phi+k_2 \hat\psi)},~~~\hat{\beta}= \beta-l\hat\psi,
\eeq
In terms of which we now have
\beq
ds^2= dx_1^2+dx_2^2+dx_3^2+dx_4^2+ \frac{k_3^2}{l^2}d\hat\beta^2,
\eeq
close to $\mu=\frac{\pi}{2}$. It is possible to show that at this point the spinor on the other hand takes the form 
\begin{align}
\hat \xi^{(5)}&= \Omega^{-1}\left(e^{\frac{i}{2l}k_3\beta}e^{-\frac{1}{2l}(k_1+k_2)\beta\gamma_{34}}\tilde{P}_++e^{-\frac{i}{2l}k_3\beta}e^{\frac{1}{2l}(k_1-k_2)\beta\gamma_{34}}\tilde{P}_-\right)e^{-\frac{\pi}{4}\gamma_{24}}e^{-\frac{\pi}{4}\gamma_{23}}e^{\frac{\pi}{4}\gamma_{45}}\eta^{(5)},\nn\\[2mm]
\Omega&= e^{\frac{1}{2}(m_1\phi+k_1\hat \psi)\gamma_{12}}e^{\frac{1}{2}(m_2\phi+ k_2 \hat {\psi})\gamma_{34}}e^{-\frac{\theta}{4}\gamma_{13}}e^{-\frac{\pi}{4}\gamma_{23}}e^{\frac{1}{4}(\theta+\pi)\gamma_{34}}e^{\frac{1}{2}X\gamma_{23}},~~~~\tilde{P}_{\pm}=\frac{1}{2}(\mathbb{I}\pm \gamma_5).
\end{align}
The point being that $\Omega$ is the spinor representation of precisely the proper local Lorentz transformation required to map the $\mu=\frac{\pi}{2}$ limit of $\hat{e}^{a}$, as defined in \eqref{eq:B5frame}, to the frame $(e')^{a}=(dx_1,dx_2,dx_3,dx_4,\frac{k_3}{l}d\hat\beta)$. Changing to this $\mathbb{R}^4\times$S$^1$ frame removes the $\Omega^{-1}$ factor from $\hat \xi^{(5)}$ and so we manifestly see that the spinor is also regular at this point as it is behaving on the covering space as a regular spinor on $\mathbb{R}^4\times$S$^1$ should, assuming $l$ is appropriately tuned as discussed earlier. We thus conclude that $\mathbb{B}^5$ preserves 4 supercharges when $l=1$, or generically just 1 when $l$ is tuned as in \eqref{eq:WCP2ls}. 

The interesting thing about the tuning of \eqref{eq:WCP2kitunings} is they allows for a spinor that is a singlet under $\partial_{\beta}$, as such it is possible to dimensionally reduce $\mathbb{B}^5$ to the un-fibred metric on any of \eqref{eq:reductiontunign} and preserve one supercharge. An example containing such an un-fibred  $\mathbb{WCP}^2$ appeared before in \cite{Macpherson:2024frt}. Notice that these tunings of $k_i$ are incompatible with $\mathbb{WCP}^2_{[1,1,1]}=\mathbb{CP}^2$, as one would expect as the un-fibred metric on $\mathbb{CP}^2$ is well known not to support a spin structure.

\subsection{$\mathbb{WCP}^3$ as a squashed $\mathbb{B}^5$ foliated over an interval}\label{sec:WCP31}
In this section we will construct an orbifolded circle fibration over a version of $\mathbb{WCP}^3_{[k_1,k_2,k_3,k_4]}$ that is based on the parametrisation of $\mathbb{CP}^3$ as a foliation of a squashed 5-sphere foliated over an interval. This too will allow for supersymmetry preserving reductions to un-fibred $\mathbb{WCP}^3$ for certain tunings of $k_i$. It is very likely that one can construct generic supersymmetric $\mathbb{WCP}^n$ in this fashion, though for $n>3$ such spaces are not likely to be relevant for string theory, so we will not peruse that here.

It is possible to write the metric on the 7-sphere as a foliation of S$^1\times$S$^5$ over an interval as
\beq
ds(\text{S}^7)= d\alpha^2+\cos^2\alpha d\chi^2+ \sin^2\alpha ds^2(\text{S}^5), \label{eq:S7met1}
\eeq
where we take the metric on the 5-sphere in \eqref{eq:S5}.
It is possible to construct a metric on $\mathbb{WCP}^3$ in much the same fashion as $\mathbb{WCP}^2$ was constructed in the previous section. One begins with the SL$(4,\mathbb{Z})$ transformation
\beq
\left(\begin{array}{c}\psi\\\phi\\ \beta\\ \chi\end{array}\right)\to \left(\begin{array}{cccc}0&-m_1&0&k_1\\
0&-m_2&0&k_2\\
1&0&0&k_3\\
0&0&-1&-k_4\end{array}\right)\left(\begin{array}{c}\psi\\\phi\\ \beta\\ \chi\end{array}\right),\label{eq:B7step1}
\eeq
where we again take $k_1m_2-k_2 m_1=1$. We then take a $\mathbb{Z}_l$ orbifold of the $\partial_{\chi}$ circle as
\beq
\chi\to \frac{1}{l}\chi.\label{eq:B7step3}
\eeq
The result is that the 7-sphere is mapped to the following 
\begin{align}
ds^2(\mathbb{B}^7)&=  d\alpha^2+ \sin^2\alpha ds^2(\mathbb{WCP}^2)+\frac{\Xi}{\Pi}\sin^2\alpha \cos^2\alpha D\beta+\frac{\Pi}{l^2}D\chi^2,~~~~\Pi=\cos^2\alpha k_4^2+\sin^2 \alpha \Xi\nn\\[2mm]
D\beta&=d\beta+{\cal B},~~~~D\chi=d\chi+{\cal C},\nn\\[2mm]
{\cal C}&= -\frac{l \Xi}{k_4\Pi}  \sin^2\alpha D\beta+\frac{l}{k_4}d\beta,~~~~{\cal B}=k_4\left(\frac{1\Delta}{k_3\Xi}\sin^2\mu D\psi\right)
\end{align}
where $(ds^2(\mathbb{WCP}^2),~D\psi,~\Xi,~\Delta)$ are defined precisely as in \eqref{eq:WCP2def} and above. We note that $\mathbb{B}^7$ is a U(1) fibration over a 6 dimensional base that is a squashed $\mathbb{B}^5$ as defined in \eqref{eq:defB5} with $l=k_4$, foliated over an interval. $\mathbb{CP}^3$ admits a parametrisation as a squashed 5-sphere foliated over an interval as
\beq
ds^2(\mathbb{CP}^3)=d\alpha^2+ \sin^2\alpha ds^2(\mathbb{CP}^2)+\sin^2\alpha \cos^2\alpha (d\beta+\eta)^2,~~~~d\eta=2 J_{\mathbb{CP}^2},
\eeq
which the base of $\mathbb{B}^7$ reduces to when $k_i=1$. We already know that $ds^2(\mathbb{WCP}^2)$ behaves as a topological $\mathbb{CP}^2$ with 3 orbifold singularites and we have that at the boundaries of $\alpha$
\beq
(\Pi,~{\cal C})\bigg\lvert_{\alpha=0}=(k_4^2,~\frac{l}{k_3}d\beta),~~~~~(\Pi,~{
\cal C})\bigg\lvert_{\alpha=\frac{\pi}{2}}=(\Xi,~-l {\cal B})
\eeq
from which it follows that the base of $\mathbb{B}^7$ is regular at $\alpha=\frac{\pi}{2}$ but tends to the origin of a cone over $\mathbb{B}^5$ as defined in \eqref{eq:defB5} with $l=k_4$ at $\alpha=0$ where ${\cal C}$ is singular. As such the base is a manifold that is topologically $\mathbb{CP}^3$ with 4 orbifold fixed points \textit{i.e.} $\mathbb{WCP}^3_{[k_1,k_2,k_3,k_4]}$, as such we identify 
\beq\label{eq:B7B5}
ds^2(\mathbb{WCP}^3)=d\alpha^2+ \sin^2\alpha ds^2(\mathbb{WCP}^2)+\frac{\Xi}{\Pi}\sin^2\alpha \cos^2\alpha D\beta^2.
\eeq
We find that the volume of the two manifolds are
\begin{align}
\text{Vol}(\mathbb{B}^7)&=\int_{\mathbb{B}^7} \text{vol}(\mathbb{B}^7)= \frac{\pi^4}{3 l},~~~~\text{Vol}(\mathbb{WCP}^3)=\int_{\mathbb{WCP}^3}\text{vol}(\mathbb{WCP}^3)\label{eq:volumeforms}\\[2mm]
&=\frac{8\pi^3}{15}\frac{k_1^2k_2^2(k_3+k_4)+k_3^2k_4^2(k_1+k_2)+ (k_1+k_2)(k_3+k_4)(k_1k_2k_3+k_2k_3k_4+k_1k_3k_4+ k_1k_2k_4)}{(k_1+k_2)(k_1+k_3)(k_1+ k_4)(k_2+k_3)(k_2+k_4)(k_3+k_4)}\nn,
\end{align}
the first of these reproducing the volume of $\text{S}^7/\mathbb{Z}_l$ and the second, while clearly being considerably more non-trivial,  reproduces $\text{Vol}(\mathbb{CP}^3)$ for $k_i=1$. Integrating the field strengths of the 3 U(1) connections over $\mathbb{WCP}^1$ we find that
\beq
\frac{1}{2\pi}\int_{\mathbb{WCP}^1} d{\cal A}=\frac{k_3}{k_1k_2},~~~\frac{1}{2\pi}\int_{\mathbb{WCP}^1,\mu=\frac{\pi}{2}} d{\cal B}=\frac{k_4}{k_1k_2},~~~-\frac{1}{2\pi}\int_{\mathbb{WCP}^1,\mu=\alpha=\frac{\pi}{2}} d{\cal C}=\frac{l}{k_1k_2}
\eeq
Finally the Euler characteristic for any 6 dimensional manifold $\mathcal{M}_6$ can be computed using the following Gauss-Bonnet formula 
\beq
\chi_E(\mathcal{M}_6)=-\frac{1}{3!(4\pi)^3}\int_{\mathcal{M}_6}\epsilon^{a_1...a_6}{\cal R}_{a_1a_2}\wedge{\cal R}_{a_3a_4}\wedge{\cal R}_{a_5a_6},\label{eq:charaterdef}
\eeq
where ${\cal R}_{ab}$ is the Ricci form. Performing this integral for $\mathbb{WCP}^3$ is very labour intensive, particularly without using a computer, but it can be computed analytically without resorting to any special tricks.  We find that 
\beq
\chi_E(\mathbb{WCP}^3)=\frac{k_1k_2k_3+k_2k_3k_4+k_1k_3k_4+k_1k_2 k_4}{k_1k_2k_3k_4}= 4-\sum_{i=1}^4\left(1-\frac{1}{k_i}\right),\label{eq:eulercharWCP3}
\eeq
yielding the expected result.

\subsubsection{Supersymmetry}
In this section we establish the supersymmetric properties of $\mathbb{ B}^7$ and the $\mathbb{WCP}^3$ it contains.\\
~~\\
We begin with the Killing spinors on the 7-sphere parameterised as in \eqref{eq:S7met1} obeying
\beq
\nabla_{m}\xi^{(7)}_{\pm}=\pm \frac{i}{2}\gamma_{m}\xi^{(7)}_{\pm}.\label{eq:7-sphereKSE}
\eeq
We take the following vielbein and basis of gamma matrices
\begin{align}
e^a&=\left(d\alpha,~\cos\alpha d\chi,~ \sin\alpha e^{a_5}\right),~~~\gamma_1=\sigma_3 \otimes \mathbb{I},\nn\\[2mm]
\gamma_2&=\sigma_1 \otimes \mathbb{I},~~~~\gamma_{a_5+2}=\sigma_2\otimes\gamma^{(5)}_{a_5}
\end{align}
where $e^{a_5}$ and $\gamma^{(5)}_{a}$ are defined as in \eqref{eq:S5viel}. With respect to this \eqref{eq:7-sphereKSE} is solved by
\begin{align}
\xi^{(7)}_{\pm}&=e^{\pm \frac{i}{2} \alpha \gamma_1}\left(e^{\frac{i}{2}\chi}(\mathbb{I}\otimes {\cal M}^{(5)}_{\mp}) Q_{\pm}+e^{-\frac{i}{2}\chi}(\mathbb{I}\otimes {\cal M}^{(5)}_{\pm}) Q_{\mp} \right)\eta^{(7)}_{\pm},~~~~Q_{\pm}=\frac{1}{2}(\mathbb{I}\pm \gamma_2),\\[2mm]
{\cal M}^{(5)}_{\pm}&=e^{\pm\frac{i}{2}\mu \gamma^{(5)}_1}\bigg(e^{\mp\frac{i}{2}\beta}(\mathbb{I}\otimes {\cal M}^{(3)}_{+}) P^{(5)}_{-}+e^{\pm\frac{i}{2}\beta}(\mathbb{I}\otimes {\cal M}^{(3)}_{-}) P^{(5)}_{+}\bigg), ~~~~ P^{(5)}_{\pm}=\frac{1}{2}(\mathbb{I}\pm \gamma^{(5)}_2)\nn
\end{align}
for $\eta^{(7)}_{\pm}$ arbitrary constant spinors and where ${\cal M}^{(3)}_{\pm}$ are defined in \eqref{eq:S3spinors}. We can generate spinors on $\mathbb{B}^7$ from the above by applying \eqref{eq:B7step1}-\eqref{eq:B7step3}, and rotating to the frame
\begin{align}
\hat e^a&=\left(d\alpha,~\sin\alpha e^{\tilde{a}}_{\mathbb{WCP}^2},~\sqrt{\frac{\Xi}{\Pi}}\sin\alpha\sin\beta D\beta,~\sqrt{\Pi}{l}D\chi\right),\nn\\[2mm]
 e^{\tilde{a}}_{\mathbb{WCP}^2}&=\left(d\mu,~\frac{1}{2}\sin\mu d\theta,~\frac{1}{2}\sin\mu \frac{\sin\theta}{\sqrt{\Delta}}d\phi,\sqrt{\frac{\Delta}{\Xi}}D\psi\right)^{\tilde{a}}.
\end{align}
With respect to this frame the spinors on ${\mathbb B}^7$ can be decomposed as
\begin{align}
\hat\xi^{(7)}_+&=\sum_{pq=\pm 1}e^{\frac{Z}{2}\gamma_{67}}e^{\frac{Y}{2}\gamma_{57}}e^{\frac{X}{2}\gamma_{47}} e^{\frac{i}{2}\alpha\gamma_1} e^{\frac{p}{2}\mu\gamma_{127}} \bigg(e^{i\Theta_{ppq}}\eta^{(7)}_{ppq}+e^{i\Theta_{(-p)pq}}e^{\frac{1}{2}\theta \gamma_{47}}\eta^{(7)}_{(-p)pq}\bigg),\nn\\[2mm]
\hat\xi^{(7)}_-&=\sum_{pq=\pm 1}e^{\frac{Z}{2}\gamma_{67}}e^{\frac{Y}{2}\gamma_{57}}e^{\frac{X}{2}\gamma_{47}} e^{-\frac{i}{2}\alpha\gamma_1} e^{\frac{p}{2}\mu\gamma_{127}} \bigg(e^{i\tilde{\Theta}_{ppq}}\eta^{(7)}_{ppq}+e^{i\tilde{\Theta}_{(-p)pq}}e^{\frac{1}{2}\theta \gamma_{47}}\eta^{(7)}_{(-p)pq}\bigg),
\end{align}
where $X,Y$ are defined as in \eqref{eq:spinspin},
\beq
\tan Z=-\frac{k_4\cot\alpha}{\sqrt{\Xi}},
\eeq
and $(\eta^{(7)}_{spq},~\tilde{\eta}^{(7)}_{spq})$ are simultaneously eigenspinors under $(i\gamma_{15},~\gamma_6,~i\gamma_{34})$ of eigenvalues $(s,~p,~q)$. This allows us to isolate in 8 independent spinors contained in each of $\xi^{(7)}_{\pm}$. The functions $(\Theta_{spq},~\Theta_{spq})$ contain the spinors dependence on the U(1) isometry directions $(\phi,~\psi,~\beta,~\chi)$, they obey
\beq
\Theta_{++\pm}=-\Theta_{--\mp},~~~\Theta_{-+\pm}=-\Theta_{+-\mp},~~~\tilde\Theta_{++\pm}=-\tilde\Theta_{--\mp},~~~\tilde\Theta_{-+\pm}=-\tilde\Theta_{+-\mp},
\eeq
such that there are only 4 independent phases in each of $\hat \xi^{(7)}_{\pm}$, they are explicitly
\begin{align}
2\Theta_{++\pm}&=\beta\mp(m_1+m_2)\phi\pm\frac{1}{l}(k_1+k_2\mp (k_3-k_4))\chi-\psi ,\nn\\[2mm]
2\Theta_{-+\pm}&=\beta\pm(m_1- m_2)\phi\mp\frac{1}{l}(k_1-k_2\mp(k_3+k_4))\chi+\psi,\\[2mm]
2\tilde{\Theta}_{++\pm}&=-\beta\pm(m_1+m_2)\phi\pm\frac{1}{l}(k_1+k_2\mp (k_3+k_4))\chi-\psi ,\nn \\[2mm]
2\tilde{\Theta}_{-+\pm}&=-\beta\pm(m_1-m_2)\phi\mp \frac{1}{l}(k_1-k_2\mp(k_3-k_4))\chi+\psi.\label{eq:B7phase}
\end{align}
We note that when $k_i=1$ then $(\tilde{\Theta}_{-++},\tilde{\Theta}_{-+\pm})$ each become independent of $\chi$, thus it is $\hat\xi^{(7)}_{-}$ that yields, in this limit, the 6 supercharges that $\mathbb{CP}^3$ preserves after reducing on $\partial_{\chi}$. For generic $(k_i,l)$ none of $(\Theta_{spq},~\tilde{\Theta}_{spq})$ are in $\frac{1}{2}\mathbb{Z}$, because of their $\chi$ dependence. They are all in  $\frac{1}{2}\mathbb{Z}$ for $l=1$, but then we only have an obnoxious parametrisations of the round 7-sphere.  For generic values of $k_i$ we can make one of $(\Theta_{++\pm},\Theta_{-+\pm},\tilde{\Theta}_{++\pm},\tilde{\Theta}_{-+\pm})$ in $\frac{1}{2}\mathbb{Z}$ with the choices 
\beq
l=k_1+k_2\mp (k_3-k_4),~~~l=k_1-k_2\mp(k_3+k_4),~~~l=k_1+k_2\mp (k_3+k_4),~~~l=k_1-k_2\mp(k_3-k_4),
\eeq
up to an irrelevant overall sign, where for each of these choices 2 supercharges of the 7-sphere are preserved. As was the case with $\mathbb{WCP}^2$ there are several tunings of $k_i\neq 1$ that make some of the phases independent of $\chi$ and so allow for generic $l$, in fact this time there are many so we will not list them all. As an example is $k_1+k_2=k_3+k_4$ which makes $\tilde{\Theta}_{+++}$ independent of $\chi$, so there are two supercharges that admit generic values of $l$ and allow for a supersymmetry preserving reduction to un-fibred $\mathbb{WCP}^3_{[k_1,k_2,k_3,k_1+k_2-k_3]}$. If we were to also take $l\in\{k_1+k_2,k_3-k_1,k_3-k_2\}$ for this tuning then 4 supercharges would be preserved on ${\mathbb B}^7$. 

A particular difference between this weighted projective space and the others is that, while the SL$(4,\mathbb{Z})$ transformation we used to generate it requires $\text{gcd}(k_1,k_2)=1$, it does not impose any particular constraints on $k_3$ and $k_4$. In particular we can choose to fix $k_3=k_1$ and $k_4=k_2$ which results in $\tilde{\Theta}_{\pm++}$ being $\chi$ independent, which means 2 things: First if we were to also take $l\in\{k_1+k_2,k_1-k_2\}$ then ${\mathbb B}^7$ preserves 6 supercharges of the 7-sphere. Second, keeping $l$ generic, this allows one to reduce to un-fibred $\mathbb{WCP}^3_{[k_1,k_2,k_1,k_2]}$ and preserve 4 supercharges. 

Of course for any of the above analysis to carry weight we need supersymmetry to be preserved globally which requires that the spinors are also regular at the boundaries of $(\theta,\mu,\alpha)$. We have checked that this is indeed the case provided that $(l,k_i)$ are tuned such that the phase in \eqref{eq:B7phase} are in $\frac{1}{2}\mathbb{Z}$ - as this analysis is very similar to that of section \ref{eq:wcp2susy} we omit the details. 

In summary we have established that there are several values of $l$  for which ${\mathbb B}^7$ preserves 2 supercharges of the 7-sphere, while there are also several tunings of $k_i$ that permit reductions to un-fibred $\mathbb{WCP}^3$, generically these preserve 2 supercharges but on  $\mathbb{WCP}^3_{[k_1,k_2,k_1,k_2]}$ 4 are actually preserved. In particular this implies that AdS$_4\times \mathbb{WCP}^3_{[k_1,k_2,k_3,k_1+k_2-k_3]}$ will generically preserve ${\cal N}=2$ superconformal symmetry, we shall return to this in section \ref{sec:AdS4wcp3}.

In the next section we will consider a different parametrisation of $\mathbb{WCP}^3$ that mirrors the parameteriation of $\mathbb{CP}^3$ as a foliation of $\mathbb{T}^{(1,1)}$ over an interval.

\subsection{$\mathbb{WCP}^3$ as a $\mathbb{T}^{(1,1)}$-like orbifold foliated over an interval}\label{sec:WCP32}
In this section we construct an alternative parametrisation of $\mathbb{WCP}^3_{[k_1,k_2,k_3,k_4]}$ which generalises the parametrisation of $\mathbb{CP}^3$ as a foliation of $\mathbb{T}^{(1,1)}$ over an interval. As we shall see, it too permits reductions to un-fibred $\mathbb{WCP}^3$ for certain tunings of $k_i$, however the flux of the connection one forms is different to that of the previous parametrisation.

Our staring point is the parametrisation of the 7-sphere as a foliation of S$^3\times$S$^3$ over an interval, its metric takes the form
\begin{align}
ds^2(\text{S}^7) &= d\mu^2+\sin^2\mu\,ds^2(\text{S}_1^3)+\cos^2\mu \,ds^2(\text{S}_2^3).\label{eq:topjointS7}
\end{align}
where we take $ds^2(\text{S}_a^3)$ to be defined as in \eqref{eqn:S3} for  $(\theta,\phi,\psi)\to(\theta_a,\phi_a,\psi_a)$ for $a=1,2$. Given section \eqref{eq:wcp1sec}, evidently the SL$(4,\mathbb{Z})$ transformation
\beq
\left(\begin{array}{c}\psi_1\\ \phi_1\\ \psi_2\\ \phi_2\end{array}\right)\to \left(\begin{array}{cccc}k_1&m_1&0&0\\k_2&m_2&0&0\\0&0&k_3&m_3\\0&0&k_4&m_4\end{array}\right) \left(\begin{array}{c}\psi_1\\\phi_1\\\psi_2\\\phi_2\end{array}\right),\label{eq:WCP32step1}
\eeq
with $k_1 m_2-k_2m_1=k_3 m_4-k_4m_3=1$ and $\text{gcd}(k_1,k_2)=\text{gcd}(k_3,k_4)=1$ will map $\text{S}_a^3\to{\mathbb B}^3_a$ with $l=1$ in \eqref{eq:topjointS7}.
The 7-sphere metric is then mapped to
\begin{align}
&ds^2= d\mu^2+\sin^2\mu\left( \frac{1}{4} ds^2(\mathbb{WCP}^1_1)+\Delta_1 (d\psi_1+{\cal A}_1)^2\right)+ \cos^2\mu\left(\frac{1}{4} ds^2(\mathbb{WCP}^1_{2})+\Delta_2 (d\psi_1+{\cal A}_1)^2\right),\nn\\[2mm]
&ds^2(\mathbb{WCP}^1_i)=\frac{1}{4}\left(d\theta_i^2+\frac{\sin^2\theta_i}{\Delta_i}d\phi_i^2\right),
\end{align}
where $(\phi_1,\phi_2,\psi_1,\psi_2)$ each have period $2\pi$ and  the connections and functions are defined as
\begin{align}
{\cal A}_1&=\bigg( \frac{m_1}{k_1}+ \frac{k_2 \sin^2\left(\frac{\theta_1}{2}\right)}{k_1 \Delta_1}\bigg)d \phi_1~~~~\Delta_1=k_1^2\cos^2\left(\frac{\theta_1}{2}\right)+k_2^2\sin^2\left(\frac{\theta_1}{2}\right)\nn\\[2mm]
{\cal A}_2&=\bigg( \frac{m_3}{k_3}+ \frac{k_4 \sin^2\left(\frac{\theta_2}{2}\right)}{k_3 \Delta_2}\bigg)d \phi_2~~~~\Delta_2=k_3^2\cos^2\left(\frac{\theta_2}{2}\right)+k_4^2\sin^2\left(\frac{\theta_2}{2}\right).\label{eq:twoWCP1}
\end{align}
We now introduce two new coordinates $(\psi,\beta)$ of period $2\pi$ through the SL$(2,\mathbb{Z})$ transformation
\beq
\left(\begin{array}{c}\psi_1\\\psi_2\end{array}\right)= \left(\begin{array}{cc}1&1\\0&1\end{array}\right)\left(\begin{array}{c}\psi\\ \beta\end{array}\right),\label{eq:WCP32step2}
\eeq  
which maps us to a circle fibration, in terms of $\partial_{\beta}$, over $\mathbb{WCP}^3$ - this mirrors how $\mathbb{CP}^3$ as a foliaton of $\mathbb{T}^{(1,1)}$ over an interval can be derived from \eqref{eq:topjointS7}. If we further take a $\mathbb{Z}_l$ orbifold of the fibre U(1) as
\beq
\beta\to \frac{1}{l}\beta,\label{eq:WCP32step3}
\eeq
the resulting metric takes the form
\begin{align}\label{eq:B7B3B3}
ds^2(\tilde{\mathbb{B}}^7)&=d\mu^2+\sin^2\mu ds^2(\mathbb{WCP}^1_1)+ \cos^2\mu ds^2(\mathbb{WCP}^1_2)+\frac{\Delta_1\Delta_2}{\tilde{\Xi}}\sin^2\mu\cos^2\mu D\psi^2+\frac{\tilde{\Xi}}{l^2} D\beta^2,\nn\\[2mm]
 \tilde{\Xi}&= \Delta_1 \sin^2\mu+\Delta_2\cos^2\mu,~~~D\psi=d\psi+\tilde{\mathcal{A}},~~~~D\beta=d\beta+\tilde{{\cal B}},
\end{align}
where the connection 1-forms are defined in terms of ${\cal A}_i$ in \eqref{eq:twoWCP1} as
\begin{align}
\tilde{\mathcal{A}}&=\mathcal{A}_1-\mathcal{A}_2,\nn\\[2mm]
\tilde{\mathcal{B}}&= \frac{l}{\tilde{\Xi}}\bigg(\Delta_1\sin^2\mu\, (d\psi+ \mathcal{A}_1)+\Delta_2\cos^2\mu \,\mathcal{A}_2\bigg),
\end{align}
and all the U(1) isometry directions have period $2\pi$. Clearly $\tilde{\mathbb{B}}^7$ is a U(1) orbifold fibration, in terms of $\partial_{\beta}$, over a 6 dimensional base that is itself a U(1) fibration, in terms of $\partial_{\psi}$, over a product of spindles. This is reminiscent of the parametrisation of $\mathbb{CP}^3$ as a foliation of $\mathbb{T}^{(1,1)}$ over an interval, which we in fact recover when $k_i=l=1$. Clearly, as it contains two spindles, this base has 4 orbifold singularities as $\mathbb{WCP}^3$ should. It is also not hard to see that at generic values of $\mu$ we have a topological $\mathbb{T}^{(1,1)}$ with orbifold fixed points at the poles of the two $\mathbb{WCP}^1$. So the base is a foliation over a generalised $\mathbb{T}^{(1,1)}$ that we will refer to as $\mathbb{WT}^{(1,1)}_{[k_1,k_2,k_3,k_4]}$, which should in general have a metric of the form\footnote{A topological $\mathbb{Y}^{p,q
}$ space with orbifold singularities has also been constructed in \cite{Ruggeri:2025kmk}, which contains another version of $\mathbb{WT}^{(1,1)}$}
\beq
ds^2(\mathbb{WT}^{(1,1)})=\lambda_1^2 ds^2(\mathbb{WCP}^1_1)+ \lambda_2^2 ds^2(\mathbb{WCP}^1_2)+\frac{\Delta_1\Delta_2}{\tilde{\Xi}}\lambda_3^2 D\psi^2,
\eeq
with $(\lambda_1,\lambda_2,\lambda_3)$ constants on this space such that the general metric on the coset SU(2)$\times$SU(2)/U(1) is recovered when $k_i=1$ - we will return to this metric in a different context in section \ref{eq:WT11ex}.
 
To confirm that we have derived a U(1) orbifold bundle over $\mathbb{WCP}^3_{[k_1,k_2,k_3,k_4]}$ what remains to confirm is that is that the foliation over $\mathbb{WT}^{(1,1)}$ behaves as it should topologically: First we observe that $\tilde{\Xi}$ is finite and non-vanishing everywhere tending to
\beq
\tilde{\Xi}(\mu=0)=\Delta_2,~~~~\tilde{\Xi}(\mu=\frac{\pi}{2})=\Delta_1,
\eeq
at the boundaries of the $\mu$ interval where $\tilde{\cal B}$ is regular. As such we have that
\beq
ds^2(\tilde{\mathbb{B}}^7/U(1))\bigg\lvert_{\mu=0}\sim d\mu^2+\mu^2\bigg[ ds^2(\mathbb{WCP}^1_1)+\Delta_1 D\psi^2\bigg]+ ds^2(\mathbb{WCP}^1_2)
\eeq
with analogous behaviour at $\mu=\frac{\pi}{2}$. Thus, at the boundaries of $\mu$, 4 directions vanish like the origin of a cone over $\mathbb{B}^3$ with the remaining two becoming a $\mathbb{WCP}^1$ of constant radius, which is exactly what one would expect for a topological $\mathbb{CP}^3$ with 4 orbifold singularities.

We thus identify
\beq
ds^2(\mathbb{WCP}^3)=d\mu^2+\sin^2\mu ds^2(\mathbb{WCP}^1_1)+ \cos^2\mu ds^2(\mathbb{WCP}^1_2)+\frac{\Delta_1\Delta_2}{\tilde{\Xi}}\sin^2\mu\cos^2\mu D\psi^2,\label{eq:WCP32}
\eeq
which parametrises $\mathbb{WCP}^3_{[k_1,k_2,k_3,k_4]}$ as a foliation of $\mathbb{WT}^{(1,1)}_{[k_1,k_2,k_3,k_4]}$ over an interval. We find that the volume of the two manifolds $\tilde{\mathbb{B}}^7$ and $\mathbb{WCP}^3$ precisely reproduce the results of  \eqref{eq:volumeforms}. Integrating the field strengths of the two U(1) connections over the two $\mathbb{WCP}^1$ factors we find that
\begin{align}
\frac{1}{2\pi}\int_{(\mathbb{WCP}^1_1,\mu=\frac{\pi}{2})}d\tilde{{\cal B}}&=-\frac{l}{2\pi}\int_{\mathbb{WCP}^1_1}d\tilde{\mathcal{A}}=\frac{l}{k_1k_2},\nn   \\[2mm]
\frac{1}{2\pi}\int_{(\mathbb{WCP}^1_2,\mu=0)}d\tilde{{\cal B}}&=\frac{l}{2\pi}\int_{\mathbb{WCP}^1_2}d\tilde{\mathcal{A}}=   \frac{l}{k_3k_4},
\end{align}
which do differ to the charges associated to the connections on the previous parametrisation of $\mathbb{WCP}^3$. Finally, through yet another lengthy and tedious calculation using \eqref{eq:charaterdef}, we have confirmed that the Euler character for \eqref{eq:WCP32} reproduces the expected result in \eqref{eq:eulercharWCP3}.

\subsubsection{Supersymmetry}\label{sec:WCP32susy}
In this section we study the supersymmetric properties of $\tilde{\mathbb{B}}^7$ and the $\mathbb{WCP}^3$ it contains.\\
~\\
We once more begin with the Killing spinors on the 7-sphere $\xi_{\pm}^{(7)}$ again obeying \eqref{eq:7-sphereKSE}. When parametrised as \eqref{eq:topjointS7} a convenient vielbien and basis of gamma matrices are the following
\begin{align}\label{eq:frameS7}
e^{a}&=(d\mu,~ \sin\mu e^{a_1},~\cos\mu e^{a_2}),~~~~\gamma_1=\sigma_2\otimes \mathbb{I}\otimes \mathbb{I}\nn\\[2mm]
\gamma_{a_1}&=\sigma_3\otimes \sigma_{a_1}\otimes \mathbb{I},~~~~\gamma_{a_2}=\sigma_1\otimes \mathbb{I}\otimes \sigma_{a_1},
\end{align}
with $e^{a_{1,2}}$ a vielbein on each 3-sphere factors as in \eqref{eq:nonnaturalframeB3}. With respect to this \eqref{eq:7-sphereKSE} is solved by
\begin{align}\label{eq:KSS7}
\xi^{(7)}_+&=e^{i\frac{\mu}{2}\gamma_1}\left(\mathbb{I}\otimes {\cal M}^{(3)}_{1-}\otimes {\cal M}^{(3)}_{2-}P_++\mathbb{I}\otimes{\cal M}^{(3)}_{1+}\otimes {\cal M}^{(3)}_{2+}P_-\right)\eta^{(7)}_+,\nn\\[2mm]
\xi^{(7)}_-&=e^{-i\frac{\mu}{2}\gamma_1}\left(\mathbb{I}\otimes {\cal M}^{(3)}_{1-}\otimes {\cal M}^{(3)}_{2+}P_++\mathbb{I}\otimes {\cal M}^{(3)}_{1+}\otimes {\cal M}^{(3)}_{2-}P_-\right)\eta^{(7)}_-,\nn\\[2mm]
P_{\pm}&=\frac{1}{2}(\mathbb{I}\pm\gamma_{1234}),
\end{align}
with $\eta^{(7)}_{\pm}$ arbitrary constant spinors and ${\cal M}^{(3)}_{{1,2}\pm}$ matrices on two 3-sphere factors defined as in \eqref{eq:S3spinors} with $(\theta,~\phi,~\psi) \to (\theta_{1,2},~\phi_{1,2},~\psi_{1,2})$.

To reach a spinor on $\tilde{\mathbb{B}}^7$ that is compatible with reductions to $\mathbb{WCP}^3$ we must apply \eqref{eq:WCP32step1}, \eqref{eq:WCP32step2} and \eqref{eq:WCP32step3} then rotate to a frame in which the fibration over $\mathbb{WCP}^3$ is manifest, we elect 
\begin{align}
\hat e^a&=\big(d\mu,~\frac{1}{2}\sin\mu  d\theta_1,~\frac{1}{2\Delta_1}\sin\mu\sin\theta_1 d\phi_1,\nn\\[2mm]
&~\frac{1}{2}\cos\mu d\theta_2,~\frac{1}{2\Delta_2}\cos\mu\sin\theta_2 d\phi_2,~\sin\mu \cos\mu\sqrt{\frac{\Delta_1\Delta_2}{\tilde{\Xi}}}D\psi,~\frac{\sqrt{\tilde{\Xi}}}{l}D\beta\big)^a.
\end{align}
With respect to this frame, the spinors on $\tilde{\mathbb{B}}^7$ take the form
\begin{align}
\hat \xi^{(7)}_+&=\sum_{p,q=\pm 1} e^{-\frac{Z}{2}\gamma_{67}}e^{-\frac{X_1}{2}\gamma_{36}}e^{-\frac{X_2}{2}\gamma_{57}} e^{i \frac{\mu}{2}\gamma_1}\left(e^{i\Theta_{+pq}}e^{-\frac{1}{2}\theta_1\gamma_{36}}e^{-\frac{1}{2}\theta_2\gamma_{57}}\eta^{(7)}_{+ pq}+e^{i\Theta_{-pq}}\eta^{(7)}_{- pq}\right),\nn\\[2mm]
\hat \xi^{(7)}_-&=\sum_{p,q=\pm 1}e^{-\frac{Z}{2}\gamma_{67}}e^{-\frac{X_1}{2}\gamma_{36}}e^{-\frac{X_2}{2}\gamma_{57}}e^{-i \frac{\mu}{2}\gamma_1}\left(e^{i\tilde{\Theta}_{+pq}}e^{-\frac{1}{2}\theta_1\gamma_{36}}\tilde{\eta}^{(7)}_{+ pq}+e^{i\tilde{\Theta}_{-pq}}e^{-\frac{1}{2}\theta_2\gamma_{57}}\tilde{\eta}^{(7)}_{- pq}\right),\label{eq:WCP32spinor}
\end{align}
where $(\eta^{(7)}_{spq},\tilde{\eta}^{(7)}_{spq})$ are eigenspinors of eigenvalues $s,p,q=\pm 1$ under respectively $\gamma_{1236}$,~~~$i\gamma_{23}$ and $i\gamma_{45}$ and 
\beq
\tan X_1=-\frac{(k_1-k_2)\sin\theta_1}{k_1+k_2+(k_1-k_2)\cos\theta_1},~~~\tan X_2=-\frac{(k_3-k_4)\sin\theta_2}{k_3+k_4+(k_3-k_4)\cos\theta_2},~~~~\tan Z=\sqrt{\frac{\Delta_1}{\Delta_2}}\tan\mu.\nn
\eeq
The various functions $(\Theta_{spq},\tilde{\Theta}_{spq})$ contain the dependence on $(\phi_1,~\phi_2,~\psi,~\beta)$, they obey
\beq
\Theta_{\pm ++}=-\Theta_{\mp--},~~~~\Theta_{\pm+-}=-\Theta_{\pm-+},~~~~\tilde{\Theta}_{\pm ++}=-\tilde{\Theta}_{\mp--},~~~~\tilde{\Theta}_{\pm+-}=-\tilde{\Theta}_{\pm-+},
\eeq
meaning that in each of $\hat \xi^{(7)}_{\pm}$ only 4 phases are independent, and each independent phase couples to 2 supercharges. Specifically we find
\begin{align}
2\Theta_{++\pm}&=\frac{1}{l}(k_1-k_2\pm(k_3-k_4))\beta+(m_1-m_2)\phi_1\mp (m_3-m_4)\phi_2+(k_1-k_2)\psi,\nn\\[2mm]
2\Theta_{-+\pm}&=-\frac{1}{l}(k_1+k_2\pm(k_3+k_4))\beta-(m_1+m_2)\phi_1\mp (m_3+m_4)\phi_2-(k_1+k_2)\psi,\nn\\[2mm]
2\tilde{\Theta}_{++\pm}&=\frac{1}{l}(k_1-k_2\mp(k_3+k_4))\beta+(m_1-m_2)\phi_1\mp (m_3+m_4)\phi_2+(k_1-k_2)\psi,\nn\\[2mm]
2\tilde{\Theta}_{-+\pm}&=-\frac{1}{l}(k_1+k_2\mp(k_3-k_4))\beta-(m_1+m_2)\phi_1\pm (m_1-m_2)\phi_2-(k_1+k_2)\psi. \label{eq:phaseWCP32}
\end{align}
Notice that this time it is 3 of $\Theta_{spq}$ that become independent of $\beta$ when $k_i=1$, so this time it is $\hat\xi^{(7)}_+$ that contains the 6 supercharges that survive the reduction of S$^7/\mathbb{Z}_l$ to $\mathbb{CP}^3$ for this tuning. Again we demand that the above phases are in $\frac{1}{2}\mathbb{Z}$ so that $\hat\xi^{(7)}_{\pm}$ have the correct periodicities as the 3 U(1)'s are traversed - this only requires tuning the coefficient of the $\beta$ terms in the above. If $l=1$ then all the phases are in $\frac{1}{2}\mathbb{Z}$, as they must be as we have an oddly parametrised  7-sphere in this limit. Again for generic values of $k_i$ we can preserve 2 supercharges by fixing
\beq
l=k_1-k_2\pm(k_3-k_4),~~~l=k_1+k_2\pm(k_3+k_4),~~~l=(k_1-k_2\mp(k_3+k_4),~~~l=k_1+k_2\mp(k_3-k_4).
\eeq
Once more there are many tunings of $k_i$ which make one of the phases independent of fibre U(1) direction, this time $\beta$. Such choices will generically lead to two supercharges being preserved on $\tilde{\mathbb{B}}^7$ that survive a reduction to $\mathbb{WCP}^3$, though 4 supercharges can be preserved on $\tilde{\mathbb{B}}^7$ for such choices when $l$ is appropriately tuned. This time in deriving $\mathbb{WCP}^3$ we had to impose $\text{gcd}(k_1,k_2)=\text{gcd}(k_3,k_4)=1$, but this still does not preclude the possibility of fixing $(k_3=k_1,k_4=k_2)$ resulting in $\theta_{-+\pm}$ being independent of $\beta$, so we again see that the phases are compatible with un-fibred $\mathbb{WCP}^3_{[k_1,k_2,k_1,k_2]}$ preserving 4 supercharges.

Thus far we have only been considering generic points along the $(\theta_1,\theta_2,\mu)$ intervals, we need to establish that $\hat\xi^{(7)}_{\pm}$ is regular at the boundaries of these intervals. As the analysis diverges from that of $\mathbb{WCP}^2$ we will be more explicit. This time the behaviour is regular at each of the boundaries of $\mu$, so there is no potential issue away from the poles of $\mathbb{WCP}_{1,2}^1$. Close to the poles of $\mathbb{WCP}^1_{1}$ we find
\begin{align}
ds^2(\mathbb{WCP}^1_{1})=(dr^2+ r^2 d\varphi_a^2),~~~~{\cal A}_1= m_a d\varphi_a,~~~~\varphi_a=\frac{1}{k_a}\phi_1~~~~a=1,2
\end{align}
which contains the only dependence on $\phi_1$, which is the only coordinate that is ill defined at this point, and where $r\in (\theta_1,\pi-\theta_1)$ at the respective poles. On the covering space of these $\mathbb{R}^2/\mathbb{Z}_{k_a}$ orbifolds, where $\varphi_a$ has period $2\pi$ we can turn off ${\cal A}_1$ via the SL$(2,\mathbb{Z})$ transformation $\psi_a=\psi+m_a \varphi_a$, yielding a regular metric on the covering space. The spinor is well defined at this point if it is regular on the covering space. We find that close to the poles the only dependence that $\hat{\xi}^{(7)}_{\pm}$ have on $\varphi_a$ is through an overall $e^{\frac{\varphi_a}{2}\gamma_{23}}$ factor, which does indeed make the spinor regular on the covering space of the poles. Regularity at the poles of $\mathbb{WCP}^1_{2}$ can be shown in much the same fashion, the only qualitative difference is that one also needs to perform an SL$(2,\mathbb{Z})$ transformation of $\beta$ to turn off the $d\phi_2$ dependence of $\tilde{\cal{ B}}$.

We conclude that some portion of the supercharges on the 7-sphere survive on $\tilde{\mathbb{ B}}^7$  whenever $(l,k_i)$ is are tuned such that the coefficient of  $\beta$ in \eqref{eq:phaseWCP32} is in $\frac{1}{2}\mathbb{Z}$. Under this constraint supersymmetry is preserved globally. 

\section{Orbifolds of near horizon geometries}\label{sec: three}
In this section we construct supersymmetry preserving orbifolds of several of the near horizon geometries that provide the canonical examples of the gravity side of the AdS/CFT correspondence. This essentially just boils down to replacing $\text{S}^n$ factors in well known solutions with the geometries $\mathbb{B}^n$ derived in the previous section. As such most of the work was already performed in the previous section, all that we need to do here is ensure that our conventions for $d=10$ supersymmetry are compatible with those used previously and present the orbifolds. This will lead to solutions containing U(1) orbifold fibrations over $\mathbb{WCP}^n$, later in sections \ref{eq:solswithroundWCPn} and \ref{eq:WT11ex}, we will utilise string dualities to generate solutions with un-fibred $\mathbb{WCP}^n$ factors from these.  
  
\subsection{D3 brane near horizon}
In this section we will construct orbifolds of the form AdS$_5\times\mathbb{B}^5$ that generalise the standard D3 brane near horizon geometry.\\
~\\
As is well known, the near horizon limit of a stack of D3 branes yields AdS$_5 \times$S$^5$ in its near horizon limit and is dual to ${\cal N}=4$ super Yang-Mills \cite{Maldacena:1997re}. One can parametrise this solution as
\begin{equation}
\frac{ds^2}{L^2} = ds^2(\text{AdS}_5) + ds^2 (\text{S}^5), \qquad e^{- \Phi} = e^{- \Phi_0} , \qquad F_5 = 4 \, L^4  e^{- \Phi_0} \left( \text{vol}(\text{AdS}_5) + \text{vol}(\text{S}^5) \right),\label{eq:AdS5S5}
\end{equation}
where the dilaton $\Phi$ and  $L$ are both  constant and AdS$_5$ and S$^5$ have unit radius. The flux quantisation requires that we fix
\beq
L^4 = 4 \pi N_3 e^{\Phi_0}, 
\eeq
 which ensures that
\begin{equation}
\frac{1}{(2 \pi )^4} \int_{\text{S}^5} F_5 =  N_3,
\end{equation}
which is the number of D3 branes.

We would like to use the results of \eqref{sec:WCP2} to construct new supersymmetric orbifolds of \eqref{eq:AdS5S5}, to this end we need information about how this geometry preserves supersymmetry\footnote{Our conventions for supersymmetry in type II supergravity follow the democratic conventions of  \cite{Tomasiello:2022dwe}, see also Appendix A of \cite{Conti:2023naw}}. For backgrounds with only a non-trivial 5-form flux and constant dilaton, supersymmety amounts to the existence of two non-trivial $d=10$ Majorana-Weyl spinors $\epsilon_{1,2}$ obeying
\begin{equation}\label{eq:10dsusyeqs}
\nabla^{(10)}_M \epsilon_1 + \frac{e^\Phi}{16} F_5 \Gamma_M \epsilon_2=0, \qquad 
\nabla^{(10)}_M \epsilon_2 - \frac{e^\Phi}{16} F_5 \Gamma_M \epsilon_1=0, \\[2mm]
\end{equation}
which imply the vanishing of the gravitino variation - the dilatino variation are trivial on such backgrounds. To make contact with \eqref{eq:AdS5S5} we decompose the $d=10$ gamma matrices on the respective AdS$_5$ and S$^5$ directions as\footnote{We work in conventions where $ \hat{\Gamma}= - \sigma_3 \otimes \mathbb{I}_4 \otimes \mathbb{I}_4$, $\gamma_{01234}= i \mathbb{I}_4$ and $\gamma_{12345} = - \mathbb{I}_4$.}
\begin{equation}\label{eq:10dgammaIIB}
\Gamma_{\mu} = L \sigma_1 \otimes \gamma_{\mu} \otimes \mathbb{I}_4, \qquad \Gamma_{a} = L \sigma_2 \otimes \mathbb{I}_4 \otimes \gamma_a,  \\[2mm]
\end{equation}
where $(\sigma_1,\sigma_2,\sigma_3)$ are the Pauli matrices.  We likewise decompose the $d=10$ spinors as
\begin{equation}\label{eq:10dspinorsAdS5S5}
\epsilon_1 = v \otimes \zeta \otimes \xi + \text{m.c.}, \qquad \epsilon_2 = i v \otimes \zeta \otimes \xi + \text{m.c.}, \qquad v = \frac{1}{\sqrt{2}}
	\begin{pmatrix}
	0 \\
	1
	\end{pmatrix},\\[2mm]
\end{equation}
where $v$ is an auxiliary 2d vector that takes care of $d=10$ chirality. $\zeta$, $\xi$ are spinors on AdS$_5$ and S$^5$ respectively and m.c is short for Majorana conjugate. Inserting the explicit value of $F_5$ from \eqref{eq:AdS5S5} in \eqref{eq:10dsusyeqs} we find that \eqref{eq:10dsusyeqs} reduce to
\begin{equation}
\nabla_{\mu} \zeta = - \frac{1}{2} \gamma_{\mu} \zeta, \qquad \text{for} \qquad M = \mu, \qquad \nabla_{a} \xi = \frac{i}{2} \gamma_{a} \xi, \qquad \text{for} \qquad M = a \\[2mm] 
\end{equation}
which are the Killing spinor equations on AdS$_5$ and S$^5$. Taking the 5-sphere metric to be \eqref{eq:S5}, the Killing spinor on S$^5$ (in the frame \eqref{eq:S5viel}) is given by \eqref{eq:S5spinor}.\\
~\\
We now replace the 5-sphere S$^5$ with $\mathbb{B}^5$ constructed in \eqref{eq:defB5}, which being locally still a 5-sphere does not effect the EOM, explicitly we have
\beq\label{eq:AdS5B5IIB}
\frac{ds^2}{L^2} = ds^2(\text{AdS}_5) + ds^2 (\mathbb{B}^5), \qquad e^{- \Phi} = e^{- \Phi_0} ,~~~
F_5  =  L^4  e^{- \Phi_0} \left( \text{vol}(\text{AdS}_5)+ \text{vol}({\mathbb{B}}^5) \right),
\eeq
with $\mathbb{B}^5$,  as defined in \eqref{eq:defB5},  a U(1) orbifold bundle over $\mathbb{WCP}^2_{[k_1,k_2,k_3]}$, that depends generically on 3 positive integers $(k_1,k_2,k_3)$ and an additional integer $l$. $\mathbb{B}^5$ is simply a rewriting of the 5-sphere when $l=1$, but for other values of $l$ it gives a non-trivial orbifolding of the 5-sphere. Locally such orbifolds support spinors like \eqref{eq:10dspinorsAdS5S5} only with $\xi$ replaced with $\hat \xi^{(5)}$ as defined in  \eqref{eq:B5spinors} with respect to the frame of \eqref{eq:B5frame}. Globally however, these orbifolds generically break supersymmetry entirely, but we have a $\frac{1}{4}$-BPS solution, which is ${\cal N}=1$ for AdS$_5$, when $l$ is tuned as one of
\beq
l= (k_1-k_2+ k_3),~l= (k_1-k_2- k_3),~ l=(k_1+k_2- k_3),~ l=(k_1+k_2+ k_3),
\eeq
which mirrors the twist and anti-twist realisations of supersymmetry preservation on the spindle. Supersymmetry is enhanced further to $\frac{1}{2}$-BPS, ${\cal N}=2$ for AdS$_5$, for certain simoltanous tunings of $(k_i,l)$. Alternatively on can keep $l\in\mathbb{Z}$ free and tune the weights to one of
\beq
k_3=k_1+k_2,~~~~k_1=k_2+k_3,~~~~ k_2=k_1+k_3,
\eeq
which results in ${\cal N}=1$ supersymmetry for generic values of $l$. Flux quantisation now leads to
\begin{equation}
\frac{1}{(2 \pi )^4} \int_{\mathbb{B}^5} F_5 = N_3,  \label{eq:D3quant} 
\end{equation}
when one tunes $\frac{ L^4 e^{-\Phi_0}}{ 4 \pi l} = N_3$.

To get the beginnings of an idea of what effect replacing S$^5$ with $\mathbb{B}^5$ has on the CFT dual we can compute the holographic Weyl anomaly on this orbifold using \cite{Henningson:1998gx}\footnote{We work in conventions such the Weyl anomaly for AdS$_5\times \text{S}^5$ is $\frac{1}{4} N_3^2$.}
\begin{equation}\label{eq:4da}
a = \frac{1}{2^6 \pi^5 } \int_{\text{M}_5} e^{3 A -2 \Phi}\text{vol}(\text{M}_5)
\end{equation}
where $e^A$ is the warp factor of unit radius AdS. For AdS$_5\times {\mathbb B}^5$ this takes the form
\begin{equation}\label{cc3}
a = \frac{1}{4}l N_3^2,\\[2mm]
\end{equation}
which does indeed depend on $l$, so diverges from the 5-sphere result.

\subsection{D1-D5 brane near horizon}
Another interesting case to consider is the D1-D5 brane near horizon geometry which takes the form, AdS$_3 \times \text{S}^3 \times \text{CY}_2$, preserves small ${\cal N}=(4,4)$ superconformal symmetry and has a compact internal space when $\mathbb{CY}_2=\mathbb{T}^4$ or $K_3$. Such backgrounds are dual to CFT on symmetric product orbifolds of the form $(\text{CY}_2)^N/S_N$ \cite{David:2002wn}. The solution can be expressed as
\begin{align}
\frac{ds^2}{L^2} & = ds^2(\text{AdS}_3) + ds^2(\text{S}^3) + \lambda^2 ds^2(\text{CY}_2), \qquad e^{- \Phi} = e^{- \Phi_0}, \nn \\[2mm]
\frac{F_3}{L^2} & = 2 e^{- \Phi_0} \left( \text{vol}(\text{AdS}_3) + \text{vol}(\text{S}^3) \right),\label{eq:D1D5sol}
\end{align}
where $(L,\lambda)$ and again the dilaton are constant. Flux quantisation requires that we tune
\beq
e^{-\Phi}L^2=N_5,~~~~4e^{-\Phi}\lambda^4 L^6 \pi^2\text{Vol}(\text{CY}_2) =N_{1},
\eeq 
where we assume a compact CY$_2$, such that 
\begin{equation}
\frac{1}{(2 \pi)^2} \int_{\text{S}^3} F_3  =N_{5}, \qquad
\frac{1}{(2 \pi)^6} \int_{\text{S}^3 \times \text{CY}_2 } \star F_3  = N_1,
\end{equation}
We can compute the leading order contribution to the central charge of the dual  CFT by means of \cite{Brown:1986nw}, leading to the formula
\begin{equation}\label{eq:2dc}
c = \frac{3}{2^4 \pi^6 } \int_{\text{M}_7} e^{A -2 \Phi}\text{vol}(\text{M}_7),
\end{equation}
which give rise to the central charge
\begin{equation}
c = 6 N_{1} N_{5},
\end{equation} 
which is consistent with the level of the small ${\cal N}=(4,4)$ superconformal algebra being $k=N_1 N_5$, to leading order in $N_1 N_5$.

To establish how supersymmetry can be preserved under the replacement S$^3\to {\mathbb{ B}}^3$, a U(1) orbifold bundle over $\mathbb{WCP}^1$, we need to know how \eqref{eq:D1D5sol} preserves supersymmetry. To this end we decompose the gamma matices on respectively AdS$_3$, S$^3_1$ and  S$^3_2$ as
\begin{equation} \label{eq:10dgammaAdS3S3CY2}
\Gamma_{\mu}  = L \, \sigma_1 \otimes \gamma_{\mu} \otimes \mathbb{I}_2 \otimes \mathbb{I}_4,~~~
\Gamma_{a}  = L \, \sigma_2 \otimes \mathbb{I}_2 \otimes \sigma_{a} \otimes \hat{\gamma}, ~~~~
\Gamma_{i}  = L \, \lambda \, \sigma_2 \otimes \mathbb{I}_2 \otimes \mathbb{I}_2 \otimes \gamma^{(4)}_{i}
\end{equation}
where we work in convensions such that the 10 and 4 dimensional chirality matrices are $\hat{\Gamma} = \sigma_3 \otimes \mathbb{I}_2 \otimes \mathbb{I}_2 \otimes \mathbb{I}_4$ and $\hat{\gamma}^{(4)} = - \gamma^{(4)}_{1234}$ respectively and $\gamma_{012}=-1$. With respect to this choice we decompose the $d=10$ Majorana-Weyl spinors as
\begin{align}
\epsilon_1  = 
	\begin{pmatrix}
	1 \\
	0
	\end{pmatrix} \otimes\bigg[ \zeta_+ \otimes \xi_+ \otimes \eta + \zeta_- \otimes \xi_- \otimes \eta \bigg]+ \text{m.c},\nn\\[2mm]
\epsilon_2 = 	\begin{pmatrix}1 \\
	0
	\end{pmatrix} \otimes \bigg[\zeta_+ \otimes \xi_+ \otimes \eta +  \zeta_- \otimes \xi_- \otimes \eta\bigg] + \text{m.c}, \label{eq:10dspinorAdS3S3CY2}
\end{align}
where $(\zeta_{\pm},\xi_{\pm},\eta)$ are distinct spinors on AdS$_3$, S$^3$ and $\text{CY}_2$ respectively.  In this case we have a supersymmetry preserving background if
\begin{equation}
\nabla^{(10)}_M \epsilon_1 + \frac{e^\Phi}{8} F_3 \Gamma_M \epsilon_2=0, ~~~~ 
\nabla^{(10)}_M \epsilon_2 + \frac{e^\Phi}{8}  F_3  \Gamma_M \epsilon_1=0,~~~~F_3 \epsilon_1=F_3 \epsilon_2 = 0\label{eq:10dsusyeqsAdS3S3CY2}.
\end{equation}
These force $\eta$ to obey $\hat{\gamma} \eta= \eta$ and be covariantly constant, as befits a spinor on a Calabi-Yau manifold, and lead to
\begin{equation}
\nabla_{\mu} \zeta_{\pm}  = {\pm} \frac{1}{2} \gamma_{\mu} \zeta_{\pm},~~~~\nabla_{a} \xi_{\pm}  = {\pm} \frac{i}{2} \gamma_{a} \xi_{\pm},\label{eq:10dsusyeqsAdS3S3CY22}
\end{equation}
making $\zeta_{\pm}$ Killing spinors on AdS$_3$ charged under distinct SL(2) subgroups of SO(2,2), and similarly for $\xi_{\pm}$ for SU(2) subgroups of SO(4) - befitting a solution with small ${\cal N}=(4,4)$ supersymmetry.

We now make the replacement S$^3\to\mathbb{B}^3$ which leads to
\begin{align}
\frac{ds^2}{L^2} & = ds^2(\text{AdS}_3) + ds^2(\mathbb{B}^3) + \lambda^2 ds^2(\text{CY}_2), \qquad e^{- \Phi} = e^{- \Phi_0}, \nn \\[2mm]
\frac{F_3}{L^2} & = 2 L^2 e^{- \Phi_0} \left( \text{vol}(\text{AdS}_3) + \text{vol}(\mathbb{B}^3) \right).\label{eq:D1D5gen1}
\end{align}
Locally this background preserves $d=10$ spinors of  the form in \eqref{eq:10dspinorAdS3S3CY2} with $\xi_{\pm}\to \xi^{(3)}_{\pm}$ as defined in section \ref{eq:WCP1susy}, but globally we are required to tune $l$ as one of\footnote{Under the assumption that we don't want to fix $k_1=k_2=1$, which results the  standard S$^3/\mathbb{Z}_l$ orbifold.}
\beq
l=k_1\pm k_2\label{eq:D1D5gen2}
\eeq
which makes only the $\xi^{(3)}_{\pm}$ spinor globally well defined, not the $\xi^{(3)}_{\mp}$ one. As such we see that orbifolding the D1-D5 near horizon as in \eqref{eq:D1D5gen1}-\eqref{eq:D1D5gen2} preserves either ${\cal N}=(4,0)$ or ${\cal N}=(0,4)$ supersymmetry. Flux quantisation now demand that we tune
\begin{equation}
e^{-\Phi}L^2=l N_5,~~~~4e^{-\Phi}\lambda^4 L^6 \pi^2\text{Vol}(\text{CY}_2) =l N_{1},
\end{equation}
which leads to the same brane charges as before. We now find that the central charge is given by
\begin{equation}
c = 6 l N_{1} N_{5},
\end{equation}
consistent with a level $k= l N_{1} N_{5}$, which explicitly depends on the weights $(k_1,k_2)$ when some supersymmetry is preserved.

\subsection{M2 brane near horizon}
We now turn our attention to the M2 brane near horizon which yields an AdS$_4 \times$ S$^7$ solution preserving maximal supersymmetry, which is ${\cal N}=8$ from the AdS$_4$ perspective. This background also admits an orbifolding as $\text{S}^7\to \text{S}^7/\mathbb{Z}_l$ which preserves ${\cal N}=6$ supersymmetry and is dual to the Chern-Simons matter theory referred to as ABJM \cite{Aharony:2008ug}, with $l$ playing the role of a Chern-Simons level. For the round 7-sphere case the metric and the flux of this solution take the form
\begin{equation}
\frac{ds^2}{L^2} = \frac{1}{4} ds^2(\text{AdS}_4) + ds^2(\text{S}^7), \qquad G = \pm \frac{3}{8} L^3 \text{vol}(\text{AdS}_4).\label{eq:AdS4S7} 
\end{equation}
Flux quantisation requires that we impose that $L^6 = 2^5 \pi^2  N_2$ such that
\begin{equation}
\frac{1}{( 2 \pi)^6} \int_{\text{S}^7} \star G  = N_2, 
\end{equation}
counting the number of M2 branes in the stack whose near horizon yields \eqref{eq:AdS4S7}.

We would like to explore the possibility of replacing S$^7$ in the above with $\tilde{{\mathbb B}}^7$, as defined in section \ref{sec:WCP32}, while preserving some supersymmetry - a similar story follows if we elect  ${\mathbb B}^7$ of section \ref{sec:WCP31}, but in the interests of brevity we will not present this. To this end we once more need information about how the starting geometry preserves supersymmetry. Supersymmetry in $d=11$ supergravity requires there to exist a non-trivial Majorana spinor $\epsilon$ such that
\begin{equation}
\nabla_M \epsilon + \frac{1}{24} \left(3 G \Gamma_M - \Gamma_M G \right) \epsilon = 0,\label{eq:deq11KSE}
\end{equation}
is obeyed. We make the following split of the $d=11$ gamma matrices in terms of AdS$_4$ and S$^7$
\begin{equation}
\Gamma_{\mu} = \frac{L}{2} \gamma_{\mu} \otimes \mathbb{I}_8, \qquad \gamma_a = L \hat{\gamma}_4 \otimes \gamma^{(7)}_a, 
\end{equation}
where $\hat{\gamma}_4 = - \gamma_{0123} $ and $i\gamma^{(7)}_{1...7} = 1$. In general a spinor on this space can decompose as
\begin{equation}\label{eq:11dspinor}
\epsilon = \zeta_+ \otimes \xi_{\pm} + \text{m.c},
\end{equation}
where $\zeta_+$ is a positive chirality spinor on AdS$_4$, $\zeta_-=\zeta_+^c$ and $\xi_{\pm}$ is a spinor on S$^7$ such that the $\pm$ subscript is correlated with $\pm$ appearing in the 4-form $G$ in \eqref{eq:AdS4S7}.
Inserting  \eqref{eq:11dspinor} and \eqref{eq:AdS4S7}  into the Killing spinor equation \eqref{eq:deq11KSE}  leads to
\begin{equation}
\nabla_{\mu} \zeta_{\pm} = \frac{1}{2} \gamma_{\mu} \zeta_{\mp}, \qquad \nabla_{\mu} \xi_{\pm} = \pm \frac{i}{2} \gamma_a \xi_{\pm}, ~~~~~  \xi_{\pm} = \pm i \xi_{\pm}^c
\end{equation}
the first two of which are the Killing spinor equations on  AdS$_4$ and S$^7$ respectively, and the second essentially ensures that we only get 8 real supercharges from the 7-sphere, giving the maximal 32 when combined with those on AdS$_4$. An explicit parametrisation of the 7-sphere is given by \eqref{eq:topjointS7}  and with respect to the frame \eqref{eq:frameS7}, the Killing spinor is given by \eqref{eq:KSS7}. 

Upon replacing S$^7$ with $\tilde{\mathbb{B}}^7$, we are mapped to the solution
\begin{equation}
\frac{ds^2}{L^2} = \frac{1}{4} ds^2(\text{AdS}_4) + ds^2(\mathbb{B}^7), \qquad G = \pm\frac{3}{8} L^3 \text{vol}(\text{AdS}_4), \label{eq:AdS4B7}
\end{equation}
where $\mathbb{B}^7$, whose precise form is given in section \ref{sec:WCP32}, is an U(1) orbifold bundle over $\mathbb{WCP}^3$ depending on 4  positive integer weights $(k_1,k_2,k_3,k_4)$ and an additional integer $l$. Flux quantisation now follows if we impose
\begin{equation}
\pm \frac{1}{( 2 \pi)^6} \int_{\tilde{\mathbb{B}}^7} \star G =  \frac{L^6}{2^5 \pi^2 l} = N_2. 
\end{equation}
Locally \eqref{eq:AdS4B7} supports Killing spinors of the form \eqref{eq:deq11KSE} with $\xi_{\pm}\to \hat\xi_{\pm}^{(7)}$ as defined in \eqref{eq:WCP32spinor}, with the constant spinors these contain constrained such that $\hat\xi^{(7)}_{\pm} = \pm i (\hat\xi^{(7)}_{\pm})^{c}$. However, as explained in section \ref{sec:WCP32susy}, global supersymmetry requires that we tune $l$, $k_i$ or both. First off for $l=1$, $\tilde{\mathbb{B}}^7$ is just the 7-sphere expressed strangely. We have a proper orbifold that generically preserves ${\cal N}=2$ supersymmetry in terms of $\frac{1}{4}$ of the supercharges contained in $\hat\xi_{+}^{(7)}$ when $l$ is tuned to one of
\beq
l=k_1-k_2+k_3-k_4,~~l=k_1-k_2-(k_3-k_4),~~~l=k_1+k_2+k_3+k_4,~~~l=k_1+k_2-(k_3+k_4).
\eeq
Supersymmetry can also be enhanced to ${\cal N}=4$ for certain simultaneous tunings of $(l,k_i)$, and ${\cal N}=6$ is achieved for $(k_3=k_1,~k_4=k_3)$ and $l=k_1\pm k_2$. Conversely, for generic values of $l$ we can preserves ${\cal N}=2$ supersymmetry by tuning the weights as one of
\beq
k_1+k_3=k_4+k_2,~~~k_1+k_4=k_2+k_3,~~~~k_1+k_2=k_3+k_4, \label{eq:k1234} 
\eeq
each of which allow one to dimensionally reduce on $\partial_{\beta}$ without breaking supersymmetry further.

For the $\hat\xi_{-}^{(7)}$ spinor we can achieve ${\cal N}=2$ supersymmetry with $l$ tuned as
\beq
l=k_1-k_2+k_3+k_4,~~~l=k_1-k_2-(k_3+k_4),~~~l=k_1+k_2+k_3-k_4,~~~l=k_1+k_2-(k_3-k_4),
\eeq
with  enhancements to ${\cal N}=4$ again possible for certain choices of $(l,k_i)$ however ${\cal N}=6$ cannot be achieved for this case for positive weights. Again we can instead elect to keep $l$ generic and instead tune the weights, this time as one of
\beq
k_2=k_1+k_3+k_4,~~~k_1=k_2+k_3+k_4,~~~k_4=k_1+k_2+k_3,~~~~k_3=k_1+k_2+k_4,\label{eq:k1234two} 
\eeq
which leads to ${\cal N}=2$ supersymmetry generically.

To get a small grip on what this is doing to the dual CFT we can compute the free energy of the CFT$_3$ thanks to \cite{Gabella:2012rc}\footnote{Note that the overall factor is slightly different from \cite{Gabella:2012rc} because they are working in conventions where AdS$_4$ has radius 2. In our case it has unit radius.}
\begin{equation}\label{eq:F3d}
{\cal{F}} = \frac{1}{2^5 \pi^6 } \int_{\text{M}_7} e^{2 A} \text{vol}(\text{M}_7)
\end{equation}
we find
\begin{equation}
{\cal{F}} = \frac{\sqrt{2} \pi }{3} l^{1/2} N_2^{3/2} .
\end{equation}
yielding explicit dependence on $l$ again.

\subsection{M5 brane near horizon}
Finally let us consider the M5 brane near horizon which is another maximally supersymmetric solution, this time of the form
\begin{align}
\frac{ds^2}{L^2} = 4 ds^2(\text{AdS}_7) + ds^2(\text{S}^4), \qquad
G = 3 L^3 \text{vol}(\text{S}^4),\label{eq:M5near}
\end{align}
where flux quantisation requires that we tune $\frac{L^3}{\pi} = N_5$, the number of M5 branes. Evidently this is not a direct product of AdS$_p$ and an odd dimensional sphere, such that we can immediately make the replacement S$^n\to \mathbb{ B}^n$ - however we can still play a similar game. If we write the metric on the 4-sphere as a foliation of S$^3$ over an interval as
\begin{equation}
ds^2(\text{S}^4) = d\alpha^2 +\sin^2\alpha  ds^2(\text{S}^3)
\end{equation}
we see that we can now replace S$^3$ in the above with $\mathbb{B}^3$, a U(1) orbifold bundle over $\mathbb{WCP}^1_{[k_1,k_2]}$, as defined in \eqref{eq:orbibundelS3}.  This leads to 
\beq
\frac{ds^2}{L^2} = 4 ds^2(\text{AdS}_7) + d\alpha^2 +\sin^2 \alpha \, ds^2(\mathbb{B}^3), \qquad
G = 3 L^3 d\alpha\wedge \sin^3\alpha\text{vol}(\mathbb{B}^3),\label{eq:M5neardef}
\eeq
This gives us a non-trivial orbifold preserving $\frac{1}{2}$ of the supercharges of \eqref{eq:M5near} when one of 
\beq
l=k_1+k_2,~~~~l=k_1-k_2
\eeq 
is elected - or one could choose to impose $k_1=k_1=1$ and keep $l$ free, but that results in a well known solution. Flux quantisation then implies
\begin{equation}
\frac{1}{(2 \pi)^3} \int_{\text{S}^4} G = \frac{L^3}{l\pi} = N_5, 
\end{equation}
while the Weyl anomaly in this case, is given by \cite{Cremonesi:2015bld}
\begin{equation}\label{eq:6da}
a = \frac{3}{2^4 7 \pi^5} \int_{\text{M}_4} e^{5 A} \text{vol}(\text{M}_4).
\end{equation}
Explicitly, we find
\begin{equation}
a = \frac{16}{7} l^2 N_5^3, 
\end{equation}
with the weights entering through $l$  when supersymmetry is preserved.

\section{AdS solutions with un-fibred $\mathbb{WCP}^n$ factors}\label{eq:solswithroundWCPn}
In this section we make use of M-theory- IIA and T- dualities to construct new  solutions with un-fibred $\mathbb{WCP}^n$ factors, mostly preserving supersymmetry. A main motivation in doing this is that they are suggestive of more general backgrounds that generically have non-trivial Romans masses which it would be interesting to construct on these foundations in the future. Another motivation is to construct the first example of AdS$_4\times \mathbb{WCP}^3$ which we achieve in section \ref{sec:AdS4wcp3} and can preserve either ${\cal N}=2$ or ${\cal N}=4$ supersymmetry, depending on how the weights are tuned.

\subsection{AdS$_7\times \mathbb{WCP}^1$}\label{sec:M5branereduction}
In this section we consider reducing \eqref{eq:M5neardef} to type IIA supergravity on the fibre U(1) of $\mathbb{B}^3$, as we shall see this leads to D6 brane sources back-reacted on a spindle, as found previously in a different context in \cite{Macpherson:2024frt}. This will break supersymmetry outside of the $k_1=k_2=1$ limit where it becomes a standard ${\cal N}=1$ reduction on S$^4/\mathbb{Z}_l$, as such there is no reason for us to tune $l$.

Following standard formulae mapping type IIA to $d=11$ supergravity we find a solution of the form
\begin{align}
ds^2 & = \frac{L^{\frac{3}{2}} \sin \alpha \sqrt{\Delta}}{\sqrt{l}} \left( 4 ds^2(\text{AdS}_7) + d \alpha^2 + \sin^2 \alpha ds^2(\mathbb{WCP}^1) \right), \qquad e^{-\Phi} = \frac{l^{\frac{3}{2}}}{L^{\frac{3}{2}} \sin^{\frac{3}{2}} \alpha \Delta^{\frac{3}{4}}}, \nn \\[2mm]
H & = - 3 L^3 \frac{\sqrt{\Delta}}{l} \sin^3 \alpha \text{vol}(\mathbb{WCP}^1) \wedge d \alpha, \qquad F_2 = \frac{2 k_1 k_2 l}{\Delta^{3/2}} \text{vol}(\mathbb{WCP}^1),
\end{align}
where $F_2=d{\cal A}$ with $(\Delta,{\cal A}, \mathbb{WCP}^1)$ defined in \eqref{eq:orbibundelS3} and below. From the flux quantisation of $G$ in $d=11$, $L^3 =l \pi N_{5}$, we have that the NS-form is quantised as 
\begin{equation}
\frac{1}{( 2 \pi)^2} \int_{\mathbb{WCP}^1 \times I_{\alpha}} H  = N_{5}, 
\end{equation}
as is standard. However, as the charge associated to the field strength of ${\cal A}$ on $\mathbb{WCP}^1$ is rational,  the charge of D6 branes inherits this property leading to
\beq
\frac{1}{2 \pi} \int_{\mathbb{WCP}^1} F_2  = \frac{l}{k_1 k_2 }. \label{eq:D6charge}
\eeq 
As $l$ is free we could of course impose that $k_1$ and $k_2$ both divide $l$ making $\frac{l}{k_1 k_2 }$ integer, however we arrive at this result by a standard string duality performed on a well defined orbifold. Thus it appears that branes backreacted on spindles may be allowed to have charges quantised in units of $\frac{1}{k_1 k_2}$ such that \eqref{eq:D6charge} is describing $l$ units of charge on the spindle - we will take this view in this and the following sections\footnote{We stress that all rational charges we find are of the form $\frac{l}{pq}$ for $l,p,q\in\mathbb{Z}$ with $l$ free. As such it is still possible to impose an integer charge by tuning $l$ - string dualities just appear to be telling us that this is not necessary, at least naively.}. It is well known that reducing AdS$_7\times \text{S}^4/\mathbb{Z}_l$ on the Hopf fibre of $\text{S}^3\subset \text{S}^4$ leads to stacks of D6 branes at the loci where the 3-sphere has vanishing radius. In this context we find something similar: About the boundaries of $\alpha$ we find that 
\beq
ds^2  = \frac{L^{\frac{3}{2}} \sqrt{\Delta}}{\sqrt{l}} \left( 4 \sqrt{r}ds^2(\text{AdS}_7) + \frac{1}{4}(dr^2 + 4 r^2 ds^2(\mathbb{WCP}^1)) \right), \qquad e^{-\Phi} = \frac{l^{\frac{3}{2}}}{L^{\frac{3}{2}}  \Delta^{\frac{3}{4}}} r^{-\frac{3}{4}},
\eeq
where $r=\alpha^2$ close to $\alpha=0$ and $r= (\pi-\alpha)^2$ close to $\alpha=\pi$, which is the behaviour close to  D6 branes extended on AdS$_7$ and back-reacted on a cone over $\mathbb{WCP}^1$. Thus the $\alpha$ interval is bounded by D6 brane singularities, albeit backreacted on $\mathbb{WCP}^1$. 

This solution is potentially interesting as, while it generically breaks supersymmetry entirely, it preserve ${\cal N}=1$ supersymmetry when $k_1=k_2=1$. Thus, like the supersymmetry breaking type IIA solution  in \cite{Macpherson:2024frt}, it is a parametric deformation of a $\frac{1}{2}$-BPS solution. There is thus some reason to hope that the background may be stable. Indeed evidence was provided in \cite{Macpherson:2024frt} that the fluctuations about such a background may be simply dressings of those of the supersymmetric case, thus inhering a spectrum above the Breitenlohner-Freedman bound from the supersymmetric case. Further it was argued in \cite{Menet:2025nbf} that similar D6 branes back-reacted on cones over spindles were also stable.

There exists an infinite family of supersymmetric AdS$_7$ solutions in massive IIA \cite{Apruzzi:2013yva,Cremonesi:2015bld}, it would be interesting to see if these too admit a supersymmetry breaking parametric deformation in terms of $\mathbb{WCP}^1$. 
  
\subsection{AdS$_5\times \mathbb{WCP}^2$}\label{sec:D3branereduction}
In \eqref{eq:AdS5B5IIB} we have a solution on AdS$_5\times\mathbb{B}^5$ in type IIB. We observe that 
\beq
ds^2(\mathbb{B}^5)=ds^2(\mathbb{WCP}^{2})+ \frac{\Xi}{l^2}\left(d\beta+{\cal B}\right)^2,
\eeq
where what appears here is defined below \eqref{eq:defB5}, and generically depends on the positive integer weights $(k_1,k_2,k_3)$. In particular $\partial_{\beta}$ is an isometry of the entire background, as such we can T-dualise on it to generate a solution in type IIA. Following the rules in \cite{Kelekci:2014ima}, we find that this leads to a solution on AdS$_5 \times \mathbb{WCP}^2 \times$ S$^1$ with non-trivial RR 4-form and NS 3-form $H=dB_2$, where specifically 
\begin{align}
ds^2 &= L^2\bigg(ds^2(\text{AdS}_5) + ds^2 (\mathbb{WCP}^2)\bigg) + \frac{l^2 }{L^2  \Xi} d \beta^2 , \qquad e^{\Phi}  = e^{\Phi_0} \frac{ l }{L \sqrt{\Xi}}, \nn\\[2mm]
B_2 & = -{\cal{B}} \wedge d\beta, \qquad F_4 = 4 e^{-\Phi_0} \frac{L^4}{ l} \sqrt{\Xi}\text{vol}(\mathbb{WCP}^2). \label{eq:AdS5IIA}
\end{align}
In order for this operation to preserve supersymmetry we need the spinors on $\mathbb{B}^5$ to be singlets with respect to $\partial_{\beta}$. As explained in section \ref{eq:wcp2susy} this does not require any restriction on $l$ but rather that we tune the weights as on of
\beq
k_3=k_1+k_2,~~~~k_1=k_2+k_3,~~~~ k_2=k_1+k_3,
\eeq
each of which preserve ${\cal N}=1$ supersymmetry, or 8 real supercharges.

It is straightforward to check that T-duality does not change the original flux quantisation condition or the Weyl anomaly behaviour, \textit{i.e.} tuning $L$ as in \eqref{eq:D3quant} and sending $N_3\to N_4$ leads to
\begin{equation}
\frac{1}{( 2 \pi)^3} \int_{\mathbb{WCP}^2}  F_4 = N_4, \qquad a = l \frac{N_4^2}{4},
\end{equation}
however the newly generated NS flux has rational charge, as was the case for D6 brane charge in the previous section, we find in this case
\beq
-\frac{1}{(2\pi)^2}\int_{\mathbb{WCP}^1\times \text{S}^1_{\beta},\mu=\frac{\pi}{2}}H_3= \frac{l}{k_1 k_2},
\eeq
which we interpret as $l$ units of NS brane charge on the spindle. Again $l$ is free, so it is possible to impose that this charge is integer. 

The IIA solution derived above is interesting in that it suggests a resolution to an apparent gap in existing AdS geometries in massive IIA. Specifically \cite{Apruzzi:2013yva} and \cite{Macpherson:2023cbl}, which preserve supersymmetry, are suggestive of a series of local solutions that take the form of AdS$_{9-2n}\times \mathbb{CP}^{n}$ foliated over an interval - in that they fill in the cases of $n=1$ and $n=3$. The case of $n=2$ does not exist in the literature but is potentially the most interesting as it contains an AdS$_5$ factor. With that said however, this would contain an un-fibred $\mathbb{CP}^2$ factor so cannot preserve supersymmetry. The solution of \eqref{eq:AdS5IIA} suggests that the true series of solutions might actually be AdS$_{9-2n}\times \mathbb{WCP}^{n}$, and preserve supersymmetry for certain tunings of the weights of the weighted projective space. It would be interesting to explore this elsewhere.

Being in massless IIA we can lift the solution of \eqref{eq:AdS5IIA} to $d=11$ to obtain AdS$_5 \times \mathbb{WCP}^2 \times \mathbb{T}^2$ of the form
\begin{align}
\frac{ds^2}{L^2} & = \frac{L^{2/3} \, \Xi^{1/3}}{l^{2/3}} e^{- \frac{2}{3} \Phi_0} \left( ds^2(\text{AdS}_5)+ ds^2 (\mathbb{WCP}^2) \right) + \frac{l^{4/3}}{ L^{10/3} \Xi^{2/3}} e^{- \frac{2}{3} \Phi_0} \left( d \beta^2 + e^{ 2 \Phi_0} d z^2 \right), \nn\\[2mm]
G & = 4 e^{- \Phi_0}\frac{L^4 }{l} \text{vol}(\mathbb{WCP}^2) + d {\cal{B}} \wedge d\beta \wedge d z.
\end{align}
In this case the flux quantisation condition yields (upon sending $N_4\to N_5$)
\begin{equation}
\frac{1}{( 2 \pi)^3} \int_{\mathbb{WCP}^2} G=N_5,~~~-\frac{1}{( 2 \pi)^3} \int_{\mathbb{WCP}^2\times \mathbb{T}^2}\star G= \frac{l}{k_1 k_2}.
\end{equation}
Upon fixing $(k_1=k_2=1,~k_3=2)$, this $d=11$ solution actually reproduces an AdS$_5\times \mathbb{T}^2\times \mathbb{WCP}^2_{[1,1,2]}$ solution from \cite{Gauntlett:2004zh}\footnote{We thank James Sparks and Jerome Gauntlett for bring this to our attention.}. Our solution provides both supersymmetric and non-supersymmetric generalisations of this. It would be interesting to establish whether this solution can be generalised to AdS$_5\times \mathbb{WCP}^2$ foliated over a generic Riemann surface.

\subsection{AdS$_4\times \mathbb{WCP}^3$}\label{sec:AdS4wcp3}
In this section we will generate an AdS$_4\times \mathbb{WCP}^3$ solution by dimensionally reducing the solution of \eqref{eq:AdS4B7} to type IIA.

Famously it is possible to reduce AdS$_4\times$S$^7$ on the Hopf fibre of the 7-sphere and arrive at a solution in type IIA preserving ${\cal N}=6$ supersymmetry \cite{Nilsson:1984bj}. This works as follows: First one parametrises S$^7$ as a circle fibration over $\mathbb{CP}^3$, then one takes a $\mathbb{Z}_l$ orbifold of the fibre resulting in
\begin{equation}
ds^2(\text{S}^7) \to ds^2(\text{S}^7/\mathbb{Z}_l) =ds^2(\mathbb{CP}^3) + \frac{1}{l^2} (d\psi+ l \, \eta)^2, \qquad d \eta = 2 J, 
\end{equation}
where $\psi$ has period $2\pi$ and $J$ is the Kahler form on $\mathbb{CP}^3$. If we now  reduce the solution of \eqref{eq:AdS4CP3}, with the above replacement, down to IIA on $\partial_{\psi}$ we obtain
\begin{equation}
\begin{split}
\frac{ds^2}{L^3/ l^2} & = \frac{1}{4} ds^2(\text{AdS}_4) + ds^2(\mathbb{CP}^3), \qquad e^{-\Phi} =  \frac{l^{3/2}}{L^{3/2}}, \\[2mm]
F_2 & = 2 \,l J , \qquad F_4 =  \frac{3}{8} L^3 \text{vol}(\text{AdS}_4). \label{eq:AdS4CP3}
\end{split}
\end{equation}
Flux quantisation requires that we tune $\frac{L^6}{2^5 \pi^2 l} = N_2$ such that 
\begin{equation}
\int_{\mathbb{CP}^1_1, \mu = \frac{\pi}{2}} F_2= 
\frac{1}{2 \pi} \int_{\mathbb{CP}^1_2, \mu = 0} F_2  =  l, ~~~~
\frac{1}{(2 \pi)^5} \int_{\mathbb{CP}^3} \star F_4  = N_2.
\end{equation}
With $L$ so tuned, the free energy in this case is given by \eqref{eq:F3d}\footnote{Since we are working in type IIA, \eqref{eq:F3d} must be weighted by a factor $e^{-2\Phi}$ since we are working in string frame and multiplied by $2 \pi$.}. We compute in this case
\begin{equation}\label{eq:FAdS4CP3}
{\cal{F}} = \frac{\sqrt{2} \pi}{3} l^{\frac{1}{2}} N_2^{\frac{3}{2}},
\end{equation}
giving rise to the well know $N^{\frac{3}{2}}$ scaling.\\
~\\
We will now construct a generalisation of \eqref{eq:AdS4CP3} starting from \eqref{eq:AdS4B7}, which is possible because
\beq
ds^2(\tilde{\mathbb{B}}^7)= ds^2(\mathbb{WCP}^3)+ \frac{\tilde{\Xi}}{l^2}(d\beta+\tilde{\cal B})^2,
\eeq
allowing us to reduce to IIA on $\partial_{\beta}$. This will preserve at least ${\cal N}=2$ supersymmetry in terms of two of the supercharges from $\xi^{(7)}_+$ or $\xi^{(7)}_-$ when the weights $(k_1,k_2,k_3,k_4)$ are tuned as one of the choices in respectively \eqref{eq:k1234} or  \eqref{eq:k1234two} - let us concretely choose the $+$ sign as this allows for a further enhancement of supersymmetry.  

Upon reducing on $\partial_{\beta}$ we find the following solution in IIA
\begin{equation}
\begin{split}
\frac{ds^2}{L^3} & = \frac{ \sqrt{\Xi}}{l} \left( \frac{1}{4} ds^2(\text{AdS}_4) + ds^2(\mathbb{WCP}^3) \right), \qquad e^{\Phi} = \frac{L^{3/2} \Xi^{3/4}}{l^{3/2}}, \\[2mm]
F_2 & = d{\tilde{\cal{B}}} , \qquad F_4 = \frac{3}{8} L^3 \text{vol}(\text{AdS}_4).
\end{split}
\end{equation}
where we remind the reader that $(ds^2(\mathbb{WCP}^3),~\tilde{\cal B})$ can be found in \eqref{eq:B7B3B3} and below.
Upon tuning $\frac{L^6}{2^5 \pi^2 l} = N_2$ the D2 brane charge is quantised in the same manor as for AdS$_4\times \mathbb{CP}^3$, \textit{i.e.}
\beq
\frac{1}{(2 \pi)^5} \int_{\mathbb{WCP}^3} \star F_4  = \frac{L^6}{2^5 \pi^2 l} = N_2.
\eeq
But similar to what we found in the previous sections, the D6 brane charge is rational, taking the form
\begin{equation}
\frac{1}{2 \pi} \int_{\mathbb{WCP}^1_1, \mu  = \frac{\pi}{2}}F_2  =  \frac{l}{k_1 k_2},~~~~
\frac{1}{2 \pi} \int_{\mathbb{WCP}^1_2, \mu  = 0 }F_2 =  \frac{l}{k_3 k_4} ,
\end{equation}
which we interprete as $l$ units of D6 brane charge on the relevant spindle - note however that these charges can be made integer if $l=k_1 k_2 k_3 k_4 \tilde{l}$ for $\tilde{l}\in \mathbb{Z}$. The free energy in this case is again of the form
\begin{equation}
{\cal{F}} = \frac{\sqrt{2} \pi}{3} l^{\frac{1}{2}} N_2^{\frac{3}{2}},
\end{equation}
matching the AdS$_4\times \mathbb{CP}^3$ case \eqref{eq:FAdS4CP3}. Let us stress that the preservation of ${\cal N}=2$ supersymmetry in this case requires that we take one of
\beq
k_1+k_3=k_4+k_2,~~~k_1+k_4=k_2+k_3,~~~~k_1+k_2=k_3+k_4,
\eeq
while there is a further enhancement to ${\cal N}=4$ if we tune $(k_3=k_1,~k_4=k_2)$. Note that while we have generated this version of AdS$_4\times \mathbb{WCP}^3$ from 
the parametrisation of \eqref{eq:B7B3B3}, the same procedure works for \eqref{eq:B7B5} with some small modifications to the finer details.

Given the obvious potential  AdS$_4/$CFT$_3$ application of the solution derived in this section it would be interesting to study it in more detail and try to nail down the CFT dual. Given the close relation of this theory to a well understood example this may well be achievable, but lies beyond the scope of this work. An interesting generalisation could also be to allow for a NS 2-form potential $B_2\sim d\tilde{{\cal B}}$ which should result in a $\mathbb{WCP}^3$ analogue of ABJ \cite{Aharony:2008gk}.

\section{AdS$_3$ with ${\mathbb B}^3\times {\mathbb B}^3$ and $\mathbb{WT}^{(1,1)}$ orbifolds}\label{eq:WT11ex}
In this section we shall return to the orbifold $\mathbb{WT}^{(1,1)}$, the orbifolded topological $\mathbb{T}^{(1,1)}$ appearing in the parametrisation of $\mathbb{WCP}^3$ in section \ref{sec:WCP32}. Here we will put it in a new context as part of a supersymmetric AdS$_3$ solution in IIA generated from an orbifold of the D1-D5 $+$ D1-D5 near horizon.\\
~\\
Our starting point is the near horizon geometry of two intersecting D1-D5 systems which is an AdS$_3\times$S$^3\times$S$^3\times$S$^1$ background preserving large ${\cal N}=(4,4)$ superconformal symmetry. This was first derived in \cite{Boonstra:1998yu} and,  after some years of confusion, was shown to have a spectrum consistent with a CFT$_2$ on symmetric product orbifold $(\text{S}^3\times\text{S}^1)^N/S_N$ in \cite{Eberhardt:2017pty}. The solution may be expressed as
\begin{align}
ds^2 & =  L^2 ds^2(\text{AdS}_3) + R_1^2 ds^2(\text{S}_1^3) + R_2^2 ds^2(\text{S}_2^3) + \lambda^2 ds^2(\text{S}^1), \qquad e^{-\Phi} = e^{-\Phi_0} ,\nn \\[2mm]
F_3 & = 2 e^{- \Phi} \left( L^2 \text{vol}(\text{AdS}_3) + R_1^2 \text{vol}(\text{S}_1^3)+ R_2^2 \text{vol}(\text{S}_2^3) \right) ,\label{eq:AdS3S3S3S1}
\end{align}
where the dilaton is a constant, we take $\text{Vol}(\text{S}^1)=2\pi$ and the radii must satisfy the following
\begin{equation}\label{eq:Radiiconstraint}
\frac{1}{L^2} = \frac{1}{R_1^2} + \frac{1}{R_2^2},
\end{equation}
for the EOM to be solved. Flux quantisation requires that we tune
\beq
e^{-\Phi}R_1^2=N_5^{(1)},~~~~e^{-\Phi}R_2^2=N_5^{(2)},~~~~e^{-\Phi} \frac{R_1^3 R_2^3\lambda}{4 L \pi}=N_1
\eeq 
such that 
\beq
\frac{1}{(2\pi)^2}\int_{\text{S}^3_a}F_3=N^{(a)}_5,~~~~~\frac{1}{(2\pi)^6}\int_{\text{S}\times\text{S}^3_1\times\text{S}^3_2}\star F_3=N_1.\label{eq:fluxquantbit}
\eeq
This leads to a central charge  of the form
\beq
c =6 \frac{N_1 N^{(1)}_5N^{(2)}_5}{N^{(1)}_5+N^{(2)}_5},
\eeq
consistent with the large ${\cal N}=4$ superconformal algebra with levels given by $(k_+= N_1 N^{(1)}_5,~k_-= N_1 N^{(2)}_5
)$.

We can construct a relatively simple orbifold out of this by parametrising the S$_a^3$ factors as in \eqref{eqn:S3} for $(\theta,\phi,\psi)\to (\theta_a,\phi_a,\psi_a)$, performing the same  SL$(4,\mathbb{Z})$ transformation on them as in \eqref{eq:WCP32step1} and then orbifolding as $\psi_a\to \frac{1}{l_a}\psi_a$. This maps S$_i^3 \to \mathbb{B}_i^3$ and the solution becomes
\begin{align}
ds^2 & =  L^2 ds^2(\text{AdS}_3) + R_1^2 ds^2(\mathbb{B}_1^3) + R_2^2 ds^2(\mathbb{B}_2^3) + \lambda^2 ds^2(\text{S}^1), \qquad e^{-\Phi} = e^{-\Phi_0} , \nn\\[2mm]
F_3 & = 2 e^{- \Phi} \left( L^2 \text{vol}(\text{AdS}_3) + R_1^2 \text{vol}(\mathbb{B}_1^3)+ R_2^2 \text{vol}(\mathbb{B}_2^3) \right) ,\label{eq:AdS3B3B3S1}
\end{align}
where  $\mathbb{B}_1^3$ depends on integers $(k_1,k_2,l_1)$ and $\mathbb{B}_2^3$ on $(k_3,k_4,l_2)$. This orbifold preserves one of the chiral  ${\cal N}=4$ factors when
\beq
{\cal N}=(4,0):~~(l_1=k_1+k_2,~l_2=k_3+k_4),~~~~~ {\cal N}=(0,4):~~(l_1=k_1-k_2,~l_2=k_3-k_4).
\eeq
To realise \eqref{eq:fluxquantbit} we must now tune
\beq
e^{-\Phi}R_1^2=l_1 N_5^{(1)},~~~~e^{-\Phi}R_2^2=l_2 N_5^{(2)},~~~~e^{-\Phi} \frac{R_1^3 R_2^3\lambda}{4 L \pi}= l_1 l_2 N_1,
\eeq 
which leads to a central charge of the form
\begin{equation}
c = 6\frac{ k_+k_-}{k_++ k_-},~~~~k_+= l_1^2l_2 N_1 N^{(5)}_1,~~~~k_+= l_1l_2^2 N_1 N^{(5)}_2\label{eq:holc}
\end{equation} 
which is still of the form required by the large ${\cal N}=4$ superconformal algebra but now explicitly depends on the weights of the two $\mathbb{WCP}^1$ factors when supersymmetry is preserved.\\
~\\ 
We can do something a bit more interesting as follows: This time we instead act on the 3-sphere isometry directions with each of \eqref{eq:WCP32step1}, \eqref{eq:WCP32step2} and \eqref{eq:WCP32step3} in sequence.  This maps us to a U(1) orbifold bundle over $\mathbb{WT}^{(1,1)}$, \textit{i.e.} now the solution takes the form
\begin{align}
ds^2&=L^2 ds^2(\text{AdS}_3)+ \lambda^2ds^2(\text{S}^1)+ ds^2(\mathbb{B}^6),~~~~e^{-\Phi}=e^{-\Phi_0},\label{eq:WT11one}\\[2mm]
F_3&= 2 e^{- \Phi} \left( L^2 \text{vol}(\text{AdS}_3)+ R_1^2\sqrt{\Delta_1}\text{vol}(\mathbb{WCP}^1)\wedge \left(d\psi+\frac{1}{l}d\beta\right)+ \frac{1}{l}R_2^2\sqrt{\Delta_2}\text{vol}(\mathbb{WCP}^2)\wedge d\beta \right)\nn.
\end{align}
The manifold $\mathbb{B}^6$ is the U(1) orbifold bundle over $\mathbb{WT}^{(1,1)}_{[k_1,k_2,k_3,k_4]}$ and takes the form
\begin{align}
ds^2(\mathbb{B}^6)&= ds^2(\widetilde{\mathbb{WT}}^{(1,1)})+ \frac{\Lambda}{l} D\beta^2,~~~~D\beta= d\beta+ \widetilde{\cal C},\nn\\[2mm]
ds^2(\widetilde{\mathbb{WT}}^{(1,1)})&= R_1^2 ds^2(\mathbb{WCP}^1)+R_2^2 ds^2(\mathbb{WCP}^2)+ \frac{\Delta_1 \Delta_2 R_1^2 R_2^2}{\Lambda}D\psi^2,~~~~D\psi=d\psi+{\cal A}_1-{\cal A}_2 \nn\\[2mm]
\Lambda&=R_1^2 \Delta_1+R_2^2\Delta_2,~~~~\widetilde{\cal C}= \frac{l}{\Lambda}\left(R_1^2\Delta_1(d\psi+{\cal A}_1)+ R_2^2 \Delta_2 {\cal A}_2\right),
\end{align}
where $(\mathbb{WCP}^1_a,~\Delta_a,~ {\cal A}_a)$ are precisely as in \eqref{eq:WCP1def} (with $(\theta,\phi)\to (\theta_a,\phi_a)$) and \eqref{eq:twoWCP1}. The integrals of the connection terms ${\cal A}_a$ over the two spindles give rise to the rational charges
\beq
\frac{1}{2\pi}\int_{\mathbb{WCP}^1}d{\cal A}_1=\frac{1}{k_1 k_2},~~~~\frac{1}{2\pi}\int_{\mathbb{WCP}^2}d{\cal A}_2=\frac{1}{k_3 k_4},
\eeq
which take the expected form. However we have been unable to construct a well defined charge from $d\widetilde{\cal C}$ unless we tune
\beq
k_3=k_1,~~~k_4=k_2,~~~m_3=m_1,~~~~m_4=m_2,\label{eq:weighttuning}
\eeq
which leads to
\beq
\frac{1}{2\pi}\int_{\Sigma_2}d\widetilde{\cal C}=\frac{l}{k_1k_2},
\eeq
where the integration cycle is $\Sigma_2=\{(\theta_1,\phi_1)~|~ \theta_2=\theta_1,~\phi_2=\phi_1\}$. We are thus considering only $\mathbb{WT}^{(1,1)}_{[k_1,k_2,k_1,k_2]}$ from this point onward. Being odd dimensional, the Euler character of this space is necessarily zero. We find that flux quantisation now requires that we tune
\beq
e^{-\Phi}R_1^2=l N_5^{(1)},~~~~e^{-\Phi}R_2^2=l N_5^{(2)},~~~~e^{-\Phi} \frac{R_1^3 R_2^3\lambda}{4 L \pi}= l N_1,\label{eq:finaltuning}
\eeq 
which leads to 
\beq
\frac{1}{(2\pi)^2}\int_{\mathbb{WCP}^1_a\times S^1_{\beta}}F_2=N^{(a)}_5,~~~~\frac{1}{(2\pi)^2}\int_{\mathbb{WCP}^1_1\times\mathbb{WCP}^1_2\times S^1_{\psi}\times S^1_{\beta}}\star F_4= N_1,
\eeq
and a central charge of the form in \eqref{eq:holc} but for levels 
\beq
k_+= l N_1 N^{(1)}_5,~~~~k_-= l N_1 N^{(2)}_5  
\eeq
\subsection{Supersymmetry}
In this section we will show that the solution of \eqref{eq:WT11one} preserves at least ${\cal N}=(2,2)$ supersymmetry when \eqref{eq:weighttuning} is imposed.

As before we require information about how \eqref{eq:AdS3S3S3S1} preserves supersymmetry to proceed. The necessary conditions for supersymmetry in this case are again \eqref{eq:10dsusyeqsAdS3S3CY2}.  We take the following $3+7$ decomposition of the gamma matrices
\beq
\Gamma_{\mu}=L \sigma_1\otimes\gamma^{(3)}_{\mu}\otimes \mathbb{I}_8 ,~~~~\Gamma_{a}=\sigma_3\otimes \mathbb{I}_2\otimes \gamma^{(7)}_a,
\eeq
such that the 10d chirality matrix is $\hat\Gamma=\sigma_3\otimes\mathbb{I}_2\otimes \mathbb{I}_8$, $i \gamma_{1234567} = \mathbb{I}_8$ and $\gamma^{(3)}_{012}=-1$. On AdS$_3$, S$^1$, S$_1^3$, and S$_2^3$ respectively. Decomposing the $d=10$ spinors as
\beq
\epsilon_1= \begin{pmatrix}
	1 \\
	0
	\end{pmatrix} \otimes\bigg[\zeta_+\otimes \chi^1_++\zeta_-\otimes \chi^1_-\bigg],~~~~\epsilon_2= \begin{pmatrix}
	1 \\
	0
	\end{pmatrix} \otimes\bigg[\zeta_+\otimes \chi^2_++\zeta_-\otimes \chi^2_-\bigg]
\eeq
where $\zeta_{\pm}$ obey the AdS$_3$ Killing spinor equation in \eqref{eq:10dsusyeqsAdS3S3CY22} and $\chi^{1,2}_{\pm}$ are spinors on the 7d internal space - each spinor is Majorana. Parametrising $ds^2(\text{S}^1)=d\varphi$ we elect the following vielbein for the internal 7d manifold
\begin{align}
e^a&=(\lambda d\varphi,~R_1e^{a_1},~R_2 e^{a_2}),\nn\\[2mm]
e^{a_{1,2}}&=\left(\frac{1}{2}d\theta_{1,2},~\frac{1}{2}\sin\theta_{1,2} (d\phi_{1,2}-d\psi_{1,2}),~d\psi_{1,2}+\sin^2\left(\frac{\theta_{1,2}}{2}\right) (d\phi_{1,2}-d\psi_{1,2})\right)^{a},\label{eqs3s3s1frame}
\end{align}
with respect to which, it is simple matter to establish that \eqref{eq:10dsusyeqsAdS3S3CY2} is solved by 
\begin{align}
\chi^1_+&=-\chi^2_+=\chi_+,~~~~\chi^1_-=\chi^2_-=\chi_-,~~~~~\chi_{\pm}={\cal M}_{\pm}\eta^{(0)}_{\pm}+\text{m.c},\nn\\[2mm]
{\cal M}_+&=e^{\frac{1}{2}(\psi_1+\phi_1)\gamma_{23}}e^{\frac{1}{2}(\psi_2+\phi_2)\gamma_{46}}P,~~~~
{\cal M}_-= e^{-\frac{\theta_1}{2}\gamma_{34}}e^{\frac{1}{2}(\phi_1-\psi_1)\gamma_{23}}e^{-\frac{\theta_2}{2}\gamma_{67}}e^{\frac{1}{2}(\phi_2-\psi_2)\gamma_{56}}P,\nn\\[2mm]
P&=\frac{L}{2}\left(\frac{1}{L}\mathbb{I}_8-\frac{i}{R_1}\gamma_{234}-\frac{i}{R_2}\gamma_{567}\right)
\end{align}
where $\eta_{\pm}^0$ are two arbitrary constant  spinors, but $P$ is a projector that reduces their number of independent components to 4. This, coupled with the fact that $\chi_{\pm}$ are Majorana,  means a total of 16 real supercharges are preserved by the background, half coupling to each of $\zeta_{\pm}$, which is consistent with ${\cal N}=(4,4)$ supersymmetry.

We wish to establish how many of these supercharges are preserved on \eqref{eq:WT11one}, we choose the vielbein
\beq
\hat e^{a}=(\lambda d\varphi,~\frac{R_1}{2}d\theta_1,~\frac{R_1\sin\theta_1}{2 \sqrt{\Delta}_1}d\phi_1,~\frac{R_2}{2}d\theta_2,~\frac{R_1\sin\theta_2}{2 \sqrt{\Delta}_2}d\phi_2,~\frac{\sqrt{\Delta_1\Delta_2}}{\sqrt{\Lambda}}D\psi,~\frac{\sqrt{\Lambda}}{l}D\beta)
\eeq
which can be generated from \eqref{eqs3s3s1frame} by first performing the coordinate transformations of \eqref{eq:WCP32step1}, \eqref{eq:WCP32step2} and \eqref{eq:WCP32step3}, fixing $(k_3=k_1,~k_4=k_2,~m_3=m_1,~m_4=m_2)$ then performing a proper Lorentz transformation. Performing the same transformations on the spinor we find that
\begin{align}
\hat\chi_+&=\sum_{p,q=\pm 1} e^{-\frac{Z}{2}}e^{-\frac{X_1}{2}}e^{-\frac{X_2}{2}}e^{i\theta^{+}_{pq}}\hat P \eta^{+}_{pq},\nn\\[2mm]
\hat\chi_-&=\sum_{p,q=\pm 1} e^{-\frac{Z}{2}}e^{-\frac{X_1}{2}}e^{-\frac{X_2}{2}}e^{i\theta^{+}_{pq}}e^{-\frac{\theta_1}{2}\gamma_{36}}e^{-\frac{\theta_2}{2}\gamma_{57}}\hat P \eta^{-}_{pq},\nn\\[2mm]
\hat P&= \frac{L}{2}\left(\frac{1}{L}\mathbb{I}_8-\frac{i}{R_1}\gamma_{236}-\frac{i}{R_2}\gamma_{457}\right)
\end{align}
where $\eta^{\pm}_{pq}$ are eigenspinors with eigenvalues $p,q=\pm 1$ under $i\gamma_{23}$ and $i\gamma_{45}$. The phases obey the identities
\begin{align}
\Theta^{\pm}_{++}=-\Theta^{\pm}_{--},~~~~\Theta^{\pm}_{+-}=-\Theta^{\pm}_{-+},
\end{align}
such that those that are independent are
\begin{align}
\Theta^{\pm}_{++}&=\mp\frac{k_1\pm k_2}{l}\beta-\frac{1}{2}(m_1\pm m_2)(\phi_1+ \phi_2)\pm\frac{1}{2}(k_1\pm k_2)\psi,\nn\\[2mm]
 \Theta^{\pm}_{+-}&=\mp\frac{1}{2}(m_1\pm m_2)(\phi_1-\phi_2)\mp\frac{1}{2}(k_1\pm k_2)\psi
\end{align}
Locally this is still consistent with 16 supercharges being preserved, but in the presence of a non-trivial orbifolding of $\partial_{\beta}$, \textit{i.e.} for $l\neq 1$, only the phases $ \Theta^{\pm}_{+-}$ are well defined for generic $l$ which yield the claimed minimal ${\cal N}=(2,2)$ supersymmmetry. This gets enhanced to either $(4,2)$ or $(2,4)$ when one additionally tunes
\beq
{\cal N}=(4,2):~l=k_1+k_2,~~~~~{\cal N}=(2,4):~l=k_1-k_2.
\eeq
One can also confirm that the spinors are well defined at the poles of the spindles, however this analysis is practically identical to the analogous part of the analysis for $\mathbb{WCP}^3$ in section \eqref{sec:WCP32susy}, so we shall omit it.

Thus we conclude that the solution of \eqref{eq:WT11one}, with \eqref{eq:weighttuning} applied, preserves ${\cal N}=(2,2)$ supersymmetry generically with a possible enhancement when $l$ is appropriately tuned. The spinors preserving ${\cal N}=(2,2)$ are also singlets with respect to $\partial_{\beta}$, meaning we can T-dualise on this direction and generate a solution with an un-fibred $\mathbb{WT}^{(1,1)}$ factor without breaking supersymmetry further. We shall do just that in the next section. 
   
\subsection{AdS$_3\times \mathbb{WT}^{(1,1)}\times\text{S}^1\times\text{S}^1$ in IIA}
As the solution of \eqref{eq:WT11one} is a U(1) orbifold bundle over $\mathbb{WT}^{(1,1)}$ in terms of $\partial_{\beta}$, we can T-dualise on this U(1) and arrive at a AdS$_3\times \mathbb{WT}^{(1,1)}\times \text{S}^1\times \text{S}^1$ solution in IIA. This will preserve ${\cal N}=(2,2)$ supersymmetry provided that we tune the weights as in \eqref{eq:weighttuning}, which is also necessary to make sense out of the charges.

Following standard T-duality formulae, for instance those found in \cite{Kelekci:2014ima}, we arrive at a solution of the form
\begin{align}
ds^2&=L^2 ds^2(\text{AdS}_3)+ \lambda^2ds^2(\text{S}^1)+ ds^2(\widetilde{\mathbb{WT}}^{(1,1)})+ \frac{l^2}{\Lambda}d\beta^2,~~~~e^{-\Phi}=e^{-\Phi_0}\frac{\sqrt{\Lambda}}{l},\label{eq:WT11two}\\[2mm]
H&=-d\beta\wedge d\widetilde{\cal C},~~~F_2=\frac{2e^{-\Phi_0}}{l}\left(R_1^2\sqrt{\Delta_1}\text{vol}(\mathbb{WCP}^1_1)+R_2^2\sqrt{\Delta_2}\text{vol}(\mathbb{WCP}^1_2)\right),\nn\\[2mm]
F_4&=2 e^{-\Phi_0}\left(L^2\text{vol}(\text{AdS}_3)+\frac{R_1^2R_2^2 \Delta_1 \Delta_2}{\Lambda}\left(\frac{1}{\sqrt{\Delta}_1}\text{vol}(\mathbb{WCP}^1_1)-\frac{1}{\sqrt{\Delta}_2}\text{vol}(\mathbb{WCP}^1_2)\right)\wedge D\psi\right)\wedge d\beta\nn.
\end{align}
In the limit $k_1=k_2=1$\footnote{Recall we have already fixed $(k_3= k_1,~k_4=k_2)$.} this solution reduces to the T-dual, on S$^1$, of an ${\cal N}=(4,2)$ solution first found in \cite{Donos:2008ug}, that appears as the IR fixed point of a flow from AdS$_5\times \mathbb{T}^{(1,1)}$ in \cite{Donos:2014eua}.

Tuning the constants again as in \eqref{eq:finaltuning}, only for $(N_1\to N_2,~N_5^{(a)}\to N_6^{(a)})$ leads to
\beq
\frac{1}{2\pi}\int_{\mathbb{WCP}^1_a}F_2= N_6^a,~~~~~-\frac{1}{(2\pi)^5}\int_{\text{S}^1\times \mathbb{WCP}^1_1\times \mathbb{WCP}^1_2\times \text{S}^1_{\psi} }\star F_4=N_2,
\eeq 
which take a standard form, however the NS 3-form gives rise to a rational charge inherited from that of $d\widetilde{\cal C}$, namely
\beq
-\frac{1}{(2\pi)^2}\int_{\Sigma_2\times \text{S}^1_{\beta}}H= \frac{l}{k_1 k_2}.
\eeq
We also have 4-form flux, but in order to define something sensible from this we need to introduce a Page charge. We define a 1-form potential such that $dC_1= F_2$ as
\beq
C_1=\frac{1}{l}e^{-\Phi_0}\left(R_1^2 \sin^2\left(\frac{\theta_1}{2}\right)d\phi_1- R_2^2\left(d\psi-\sin^2\left(\frac{\theta_2}{2}\right)d\phi_2\right)\right).
\eeq
then $\hat F_4= F_4+H\wedge C_1$ is closed and so suitable for defining a charge. We find
\beq
-\frac{1}{(2\pi)^3}\int_{\mathbb{WCP}^1_2\times S^1_{\psi}\times S^1_{\beta}}\hat F_4=N_4= l N_6^{(2)}.
\eeq
The central charge of this solution is the same as it was in IIB only with $(N_1\to N_2,~N_5^{(a)}\to N_6^{(a)})$.

Given that the solution of \eqref{eq:WT11two}, up to T-duality, is a parametric deformation of the IR fixed point of a flow from AdS$_5\times \mathbb{T}^{(1,1)}$ \cite{Donos:2014eua}, it is natural to wonder if AdS$_5\times \mathbb{WT}^{(1,1)}$ also exists. Of course $\mathbb{WT}^{(1,1)}$ is not a Sasaki-Einstein manifold, so such a solution would presumably need to contain additional warp factors and a non-constant dilaton that depend on the directions of $\mathbb{WCP}^1_{a}$. While it would be interesting to construct such a solution, it should be inaccessible via string dualities and as such would require a focused effort to construct from first principles. We leave this potentially interesting avenue to be explored elsewhere.  
  
\section{Conclusions}\label{sec:conclusions}
In this work we have constructed U(1) orbifold bundles over the weighted projective spaces $\mathbb{WCP}^n$ for $n=2,3$. These are higher dimensional analogues of the spindle which is $\mathbb{WCP}^1$. We have constructed these spaces starting from round $(2n+1)$-spheres, studied their geometric and supersymmetry properties and established that, for certain tunings of their integer weights, they allow a Kaluza-Klein reduction to un-fibred $\mathbb{WCP}^n$.

We leverage these results to construct non-trivial orbifolds of well known AdS solutions of importance to the AdS/CFT correspondence. These orbifold are expressed in terms of the U(1) orbifold bundles over $\mathbb{WCP}^n$ and preserve some portion of the original supersymmetry. Specifically we construct ${\cal N}=1$ preserving orbifolds of AdS$_7\times$S$^4$,  generically ${\cal N}=1$ preserving orbifolds of AdS$_5\times$S$^5$, generically ${\cal N}=2$ preserving orbifolds of AdS$_4\times \text{S}^7$ and  ${\cal N}=(4,0)$ preserving orbifolds of AdS$_3\times$S$^3\times \text{CY}_2$. The AdS$_5$ and AdS$_4$ examples can also experience enhancements to their supersymmetry when the integer weights of the $\mathbb{WCP}^n$ factors they contain are tuned appropriately.

We have also made use of supergravity dualities, specifically T-duality of type II and type IIA/ M-theory duality, to derive examples of AdS solutions containing un-fibred $\mathbb{WCP}^n$ factors. These examples include a two parameter supersymmetry breaking deformation of the IIA reduction of AdS$_7\times\text{S}^4$, a supersymmetric AdS$_5\times \mathbb{WCP}^2\times \text{S}^1$ solution in IIA and, perhaps most interestingly, a supersymmetric AdS$_4\times \mathbb{WCP}^3$ solution that yields a 3 parameter deformation of AdS$_4\times \mathbb{CP}^3$.  Beyond $\mathbb{WCP}^n$,  we also constructed a generically ${\cal N}=(2,2)$ supersymmetric AdS$_3\times \text{S}^1\times \text{S}^1\times \mathbb{WT}^{(1,1)}$ solution in type IIA, where   $\mathbb{WT}^{(1,1)}$ is topologically $\mathbb{T}^{(1,1)}$ with 4 orbifold singularities.

Something unusual about the solutions we have found with un-fibred $\mathbb{WCP}^n$ factors is that they lead to certain brane charges being rational, rather than integer. At first sight this may seem perverse, however it is a property they inherit from the rational charge of the field strength of the connection on the spindle.  This is well defined, and the operations that map this to the charge of a D6 brane under a reduction to IIA or an NS5 brane under T-duality are likewise well defined. 
Thus it seems reasonable to at least entertain the idea that rational charges for branes backreacted on weighted projective spaces may be possible - this would be interesting to explore in more detail. However, if one wishes to take a more conservative view, note that each of the rational charges we find is of the form $\frac{l}{pq}$ for $l,p,q\in\mathbb{Z}$, with $l$ free, so it is still possible to impose a standard Dirac quantisation condition for these branes by tuning $l$.   

It is worth stressing that while we have constructed supersymmetric solutions containing $\mathbb{WCP}^n$ factors for $n=1,2,3$, these appear in a different context to the spindle in most of the literature. Specifically, starting in \cite{Ferrero:2020laf}, most of the focus has been on the geometries describing branes that wrap a spindle and their dual compactified CFT description. The geometries we construct are essentially the near horizon limits of branes that are backreacted on orbifolds containing $\mathbb{WCP}^n$ factors, which is more similar to \cite{Macpherson:2024frt}. It would thus be interesting to see whether the full monopole solutions that contain these orbifolds can also be constructed, similar to how \cite{Gauntlett:1997pk} realises the $\text{S}^7/\mathbb{Z}_k$ orbifold of ABJM in \cite{Aharony:2008ug}.

We elected to construct orbifolds of geometries with well understood CFT duals in this work. We did make a minimal attempt to understand the effect that our orbifolds have on the CFT duals by computing quantities such as the central charge, free energy and Wely anomaly. However there is a lot more that needs to be done in this direction to get a full picture. Our hope is that our work will inspire a more focused effort in this direction.\\
~\\
Having established that $\mathbb{WCP}^n$ for $n>1$ are compatible with supersymmetry there are many interesting future directions to explore, let us mention just a few
\begin{itemize}
\item The procedure we apply to construct orbifolds containing $\mathbb{WCP}^n$ factors in this work is quite broadly applicable. Indeed it can essentially be applied to any background containing a $(2n+1)$-sphere for $n>0$. It may be fruitful to explore what else can be generated through orbifolds and string dualities following the path taken in this work.
\item Beyond possessing a potentially interesting CFT dual, the  $\text{AdS}_4\times \mathbb{WCP}^3$ solution we have found is interesting because broad classes of AdS$_2$ and AdS$_3$ solutions containing $\mathbb{CP}^3$ factors also exist \cite{Macpherson:2023cbl,Lozano:2024idt,Lozano:2025ief,Conti:2025djz}. Perhaps these too admit a generalisation with $\mathbb{CP}^3\to \mathbb{WCP}^3$.
\item Several of the geometries we have derived in this work are suggestive of broader classes of interesting solutions. One of the most interesting of these is a potential class of superymmetric solutions with AdS$_5\times \mathbb{WCP}^2$ foliated over an interval in massive IIA, which the solution of section \ref{sec:D3branereduction} suggests - see the discussion therein.
\item Finally, while $\mathbb{WCP}^3$ is likely too large, it would be very interesting to realise $\mathbb{WCP}^2$ in some wrapped brane scenario more similar to the bulk of the literature on spindles. A likely fruitful avenue to explore is M5 branes wrapping $\mathbb{WCP}^2$ dual to $d=6$ CFTs compactified on the same space.
\end{itemize}
We hope that we or others will report on such interesting lines of enquiry in the future.

\section*{Acknowledgements}
We thank Paul Merrikin for collaboration during the early stages of this project and  Christopher Couzens for some enlightening discussions. We also thank Jerome Gauntlett and James Sparks for useful comments. NTM and AC acknowledge support from grants from the Spanish government MCIU-22-PID2021-123021NB-I00, MCIU-25-PID2024-161500NB-I00 and principality of Asturias SV-PA-21-AYUD/2021/52177. The work of NTM is also funded by the Ram\'on y Cajal fellowship RYC2021-033794-I and AC by the Severo Ochoa fellowship PA-23-BP22-019. The work of NTM was also supported in part by Simons Foundation award number 1023171-RC. AC thanks Imperial College London for its kind hospitality while this work was being completed.

\end{document}